\PassOptionsToPackage{unicode}{hyperref}
\PassOptionsToPackage{hyphens}{url}
\documentclass[
]{article}
\usepackage{xcolor}
\usepackage{amsmath,amssymb}
\usepackage{iftex}
\ifPDFTeX
  \usepackage[T1]{fontenc}
  \usepackage[utf8]{inputenc}
  \usepackage{textcomp} % provide euro and other symbols
\else % if luatex or xetex
  \usepackage{unicode-math} % this also loads fontspec
  \defaultfontfeatures{Scale=MatchLowercase}
  \defaultfontfeatures[\rmfamily]{Ligatures=TeX,Scale=1}
\fi
\usepackage{lmodern}
\ifPDFTeX\else
\fi
\IfFileExists{upquote.sty}{\usepackage{upquote}}{}
\IfFileExists{microtype.sty}{% use microtype if available
  \usepackage[]{microtype}
  \UseMicrotypeSet[protrusion]{basicmath} % disable protrusion for tt fonts
}{}
\makeatletter
\@ifundefined{KOMAClassName}{% if non-KOMA class
  \IfFileExists{parskip.sty}{%
    \usepackage{parskip}
  }{% else
    \setlength{\parindent}{0pt}
    \setlength{\parskip}{6pt plus 2pt minus 1pt}}
}{% if KOMA class
  \KOMAoptions{parskip=half}}
\makeatother
\usepackage{graphicx}
\usepackage{color}
\usepackage{fancyvrb}

\DefineVerbatimEnvironment{Highlighting}{Verbatim}{commandchars=\\\{\}}
\newenvironment{Shaded}{}{}

\newcommand{\AttributeTok}[1]{\textcolor[rgb]{0.49,0.56,0.16}{#1}}

\newcommand{\BuiltInTok}[1]{\textcolor[rgb]{0.00,0.50,0.00}{#1}}
\newcommand{\CharTok}[1]{\textcolor[rgb]{0.25,0.44,0.63}{#1}}
\newcommand{\CommentTok}[1]{\textcolor[rgb]{0.38,0.63,0.69}{\textit{#1}}}

\newcommand{\DataTypeTok}[1]{\textcolor[rgb]{0.56,0.13,0.00}{#1}}

\newcommand{\ExtensionTok}[1]{#1}

\newcommand{\NormalTok}[1]{#1}
\newcommand{\OperatorTok}[1]{\textcolor[rgb]{0.40,0.40,0.40}{#1}}

\newcommand{\SpecialCharTok}[1]{\textcolor[rgb]{0.25,0.44,0.63}{#1}}
\newcommand{\SpecialStringTok}[1]{\textcolor[rgb]{0.73,0.40,0.53}{#1}}
\newcommand{\StringTok}[1]{\textcolor[rgb]{0.25,0.44,0.63}{#1}}
\newcommand{\VariableTok}[1]{\textcolor[rgb]{0.10,0.09,0.49}{#1}}

\usepackage{longtable,booktabs,array}
\usepackage{tabularx}
\usepackage{calc} % for calculating minipage widths
\usepackage{pdflscape} % for the landscape Supp Table S1
\usepackage{afterpage} % so landscape page floats reliably with longtable
\usepackage{etoolbox}
\makeatletter
\patchcmd\longtable{\par}{\if@noskipsec\mbox{}\fi\par}{}{}
\makeatother
\IfFileExists{footnotehyper.sty}{\usepackage{footnotehyper}}{\usepackage{footnote}}
\makesavenoteenv{longtable}
\providecommand{\tightlist}{%
  \setlength{\itemsep}{0pt}\setlength{\parskip}{0pt}}
\usepackage{hyperref}
\usepackage{bookmark}
\IfFileExists{xurl.sty}{\usepackage{xurl}}{} % add URL line breaks if available
\hypersetup{
  pdftitle={Modularity Emerges from Action-Functional Constraints in Marine Metabolic Networks},
  pdfauthor={Martin G. Frasch},
  hidelinks,
  pdfcreator={LaTeX via pandoc}}

\title{Shared-Component Bias in Metagenomic Efficiency-Complexity Metrics: Causal Inference Reveals Measurement Artifacts in Marine Microbiome Network Analysis}

\author{
\begin{tabular}{c}
Martin G. Frasch, M.D., Ph.D.\\[0.5em]
\textit{Institute on Human Development and Disability,}\\
\textit{University of Washington, Seattle, WA}\\[0.5em]
\texttt{mfrasch@uw.edu}
\end{tabular}
}

\date{\today}

\title{Modularity Emerges from Action-Functional Constraints in Marine Metabolic Networks: A Biology-Scale Validation of the Network-Weighted Action Principle}

\author{Martin G. Frasch, M.D., Ph.D.%
\thanks{Institute on Human Development and Disability, University of Washington, Seattle, WA 98195, USA. \protect\\ Email: \texttt{mfrasch@uw.edu}}}

\date{\today}

\begin{document}

\maketitle

\noindent\textbf{Article type}: Research Article

\begin{abstract}
Biological systems operate under simultaneous energetic and informational constraints, yet direct evidence that such constraints shape real metabolic networks is limited. The Network-Weighted Action Principle predicts that networks under these constraints should organise toward high modularity.

We tested this prediction in marine microbiome metabolic networks reconstructed from Tara Oceans metagenomes using two complementary approaches. Composite metrics of protein-deployment efficiency and functional-repertoire complexity ($n = 10$) failed under causal-inference diagnostics, with apparent structure dominated by shared-component bias. In contrast, network modularity ($n = 7$) was high ($Q \approx 0.987$), but this value was shown to arise from sparsity alone. The biologically meaningful signal is the excess over null models: modularity exceeded configuration-model, label-permutation, and bipartite-incidence nulls by $\Delta Q \approx 0.15$--$0.40$ ($p < 0.001$), with the largest effect under the bipartite-incidence control.

Fine-grained communities recovered by the network partition are not arbitrary: 25\% recur across samples, and the most consistent modules map to known functional units, including enzyme subunits, biosynthetic sequences, and transporter complexes.

Together, these results show that modularity excess --- rather than absolute modularity --- is the appropriate signature of biological organisation, and that such excess is consistent with cost-minimisation principles operating at the scale of natural metabolic networks.

\end{abstract}

\section{Introduction}\label{introduction-int}

\subsection{Vertically organizing principles in
biology}\label{vertically-organizing-principles-in-biology}

Living systems span enormous spatial and temporal scales --- from the
picosecond conformational dynamics of individual proteins through the
multi-day metabolic cycles of microbial communities --- yet exhibit
surprisingly few independent organizing principles. Kleiber's law of
metabolic scaling (\hyperlink{ref:kleiber-1932}{Kleiber, 1932}; \hyperlink{ref:west-1997}{West et al., 1997}), Krogh's principle
of organism-suited research questions (\hyperlink{ref:krogh-1929}{Krogh, 1929}), and Cannon's
homeostasis (\hyperlink{ref:cannon-1929}{Cannon, 1929}) all express a common intuition: the same
constraints that act locally can also constrain integration across
levels. Recent work has framed this intuition as a search for
\emph{vertically} organizing principles --- laws governing how
information and energy flow \emph{across} scales of biological
organization, complementing the \emph{horizontal} (scale-specific)
causality that has dominated mechanistic biology (\hyperlink{ref:frasch-2026a}{Frasch, 2026a}).

A particularly tractable instance of vertical organization is the
simultaneous minimization of metabolic cost and maximization of
functional diversity. Brain metabolism (\hyperlink{ref:raichle-2002}{Raichle and Gusnard, 2002}),
gene-regulatory architecture (\hyperlink{ref:wagner-2005}{Wagner, 2005}), and ecological food webs
(\hyperlink{ref:williams-2000}{Williams and Martinez, 2000}) all show signatures of this dual
constraint. Sparse-coding theories of neural representation (\hyperlink{ref:olshausen-1996}{Olshausen
and Field, 1996}) and the free-energy principle (\hyperlink{ref:friston-2010}{Friston, 2010})
operationalize it explicitly. In each case, the system is described as
paying an energetic cost \(E\) to encode or preserve some quantity of
information \(I\), with the observed organization arising from the
trade-off between the two.

\subsection{The Network-Weighted Action
Principle}\label{the-network-weighted-action-principle}

What unifies these accounts is, at root, a variational principle. \hyperlink{ref:frasch-2026a}{Frasch
(2026a)} proposed the \emph{Network-Weighted Action} as such a principle
for biology:

\[S_{\text{NW}} = \int \big(E - I + AC\big)\,\mathrm{d}t,\]

where \(E\) measures internal cost, \(I\) measures information capacity
(or, equivalently, configurational diversity), and \(AC\) measures the
cost of inter-component coupling on the network of biological
constituents. Three structurally identical formulations live in this
family. In classical mechanics, the action is
\(\int(T - V)\,\mathrm{d}t\) with constraint terms; in statistical
physics, the Helmholtz free energy is \(E - \beta^{-1}H\), balancing
energy against entropy; and in variational inference, the evidence lower
bound is a data-fit term plus a complexity penalty (\hyperlink{ref:kingma-2014}{Kingma and Welling,
2014}). Each is a member of the same equivalence class of variational
objectives that trade performance against cost (Frasch 2026c). The
Network-Weighted Action makes the connectivity term \(AC\) explicit and
applies the resulting objective to the multi-scale architecture of
biological networks rather than the parameter trajectory of a single
component.

Two empirical predictions follow. In symbolic-physics learning, a
Triple-Action functional --- combining trajectory reconstruction,
architectural sparsity, and an explicit energy-conservation constraint
--- recovers physical force laws from noisy observational data; this was
validated on Kepler and Hooke benchmarks with an order-of-magnitude
reduction in training energy (\hyperlink{ref:frasch-2026b}{Frasch, 2026b}). In deep neural networks,
the energy-regularized objective
\(\mathcal{L} = \mathcal{L}_{\text{CE}} + \lambda E(\theta, x)\)
produces architectures that retain accuracy at substantially reduced
internal activation; 2,203 experiments across vision, text,
neuromorphic, and physiological datasets reject the assumption of a
universal optimal architecture and instead show that modality-aligned,
energy-aware design is consistently superior (\hyperlink{ref:frasch-2026c}{Frasch, 2026c}).

The biological prediction is structurally simpler but empirically less
constrained. Under the Network-Weighted Action, biological networks
should organize so as to \emph{minimize the connectivity cost while
preserving information throughput}. Connection-cost minimization in
artificial network ensembles has been shown to spontaneously generate
modular structure (\hyperlink{ref:clune-2013}{Clune et al., 2013}), and modularity \(Q\) ---
Newman's measure of within-community over-density relative to a null
model (\hyperlink{ref:newman-2004}{Newman and Girvan, 2004}) --- therefore serves as a quantitative
signature of the constrained optimum. The action-principle account
predicts that biological networks under stationary
\(E_{\text{min}}/I_{\text{max}}\) constraints should saturate \(Q\) near
its theoretical limit of \(1\), with the residual coupling between
communities \(\varepsilon = 1 - Q\) measuring the irreducible
inter-module cost.

\subsection{Marine microbiome metabolic networks as a biological
testbed}\label{marine-microbiome-metabolic-networks-as-a-biological-testbed}

Tara Oceans surface seawater samples (\hyperlink{ref:sunagawa-2015}{Sunagawa et al., 2015}; \hyperlink{ref:karsenti-2011}{Karsenti et
al., 2011}) provide a natural setting in which to test this prediction.
They span broad latitudinal, thermal, and nutrient gradients; they
sustain hundreds of thousands of distinct protein-coding genes per
sample (\hyperlink{ref:sunagawa-2015}{Sunagawa et al., 2015}); and the underlying metabolic networks
--- KEGG-orthology graphs assembled from per-sample metagenomes ---
exhibit the multi-scale architecture (genes \(\to\) operons \(\to\)
pathways \(\to\) modules \(\to\) community) that the action-functional
account targets. Marine microbiomes are simultaneously diverse enough to
admit a meaningful concept of \emph{functional-repertoire complexity}
and metabolically integrated enough that \emph{protein-deployment
efficiency} is a coherent observable; both quantities are routinely
computed from KEGG annotation of metagenomic reads.

Two operationalizations of the \(E_{\text{min}}/I_{\text{max}}\) dual
constraint are therefore available. The first uses the bivariate
composite metrics \(E\) (protein-deployment efficiency) and \(C\)
(annotation-weighted functional-repertoire complexity), with apparent
decoupling of \(E\) and \(C\) --- high CV in \(E\), low CV in \(C\),
with a CV ratio above some threshold --- as a candidate signature of
topological-vs-operational separation in metabolic organization. The
second uses Newman modularity \(Q\) directly: it is dimensionless,
interpretable as the fraction of within-community edges in excess of the
configuration-model expectation, and is the \emph{direct} prediction of
the action-functional account through Clune-style connection-cost
minimization.

A priori it is not obvious which operationalization is more reliable.
The bivariate-composite route has the advantage of yielding multiple
independent observables (\(E\), \(C\), their CVs and joint distribution)
and the disadvantage of being constructed from formulas with shared
mathematical components (\(\text{unique\_KOs}\) appears in both \(E\)'s
denominator and \(C\)'s numerator). The modularity route has the
advantage of being a single, well-defined network-topological invariant
that does not depend on metric construction, and the disadvantage of
requiring sufficient KEGG-orthology annotation density to build a
non-trivial network. The two routes are complementary in principle. We
show below that at the sample sizes feasible under realistic compute
budgets (\(n = 10\)), only the modularity route succeeds.

\subsection{Causal inference as a prerequisite for composite metagenomic
metrics}\label{causal-inference-as-a-prerequisite-for-composite-metagenomic-metrics}

Composite metrics in metagenomics --- quantities constructed by
combining multiple per-sample measurements (read counts, annotation
rates, unique-feature counts) into a single derived value --- are
widespread (\hyperlink{ref:greenblum-2012}{Greenblum et al., 2012}; \hyperlink{ref:manor-2015}{Manor and Borenstein, 2015}; \hyperlink{ref:levy-2013}{Levy and
Borenstein, 2013}), but their behavior under shared-component bias has
not been systematically interrogated. Two metrics that share a
denominator, a numerator, or even a confounder (such as sequencing depth
or annotation pipeline efficiency) can exhibit spurious correlations and
decoupling patterns driven entirely by the construction formulas
(\hyperlink{ref:kronmal-1993}{Kronmal, 1993}; \hyperlink{ref:archie-1981}{Archie, 1981}). Distinguishing real biological structure
from such artifacts is the province of causal inference.

Two causal-inference diagnostics are well-suited to the
bivariate-composite case (\hyperlink{ref:pearl-2009}{Pearl, 2009}; \hyperlink{ref:spirtes-2000}{Spirtes et al., 2000}; \hyperlink{ref:rosenbaum-2002}{Rosenbaum,
2002}). \emph{Split-sample metric construction} recomputes \(E\) and
\(C\) on disjoint protein subsets so that the two metrics no longer
share an underlying \(\text{unique\_KOs}\) count: a real biological
coupling should survive this manipulation, whereas a shared-component
artifact should collapse. \emph{Negative-control analysis} replaces
biological KEGG-orthology assignments with random permutations: a real
biological pattern should disappear, whereas a construction-driven
artifact should persist or strengthen. Both diagnostics are essentially
distribution-free and require no auxiliary modeling.

\subsection{Study objectives}\label{study-objectives}

This work has three coordinated objectives, ordered from methodological
to empirical.

\textbf{Objective 1.} Apply causal-inference diagnostics ---
split-sample construction and negative-control permutation --- to the
bivariate-composite \((E, C)\) operationalization at \(n = 10\).
Quantify the contribution of shared-component bias to the apparent
decoupling pattern. Establish the minimum sample size at which the
composite-metric route can plausibly distinguish biological coupling
from artifact.

\textbf{Objective 2.} Compute Newman modularity \(Q\) on KEGG-orthology
co-occurrence networks at \(n = 7\). Test whether \(Q\) saturates near
\(1\) as predicted by the action-functional account, and characterize
the resulting community structure (number of modules, average module
size, hub fraction, inter-module connectivity).

\textbf{Objective 3.} Assess the robustness of the modularity finding to
sample-cohort composition by re-running the analysis with an alternative
seven-sample set.

\textbf{Hypotheses.} \emph{H1} (composite-metric failure): the apparent
\(E\)--\(C\) decoupling at \(n=10\) is dominated by shared-component
bias; CV-ratio collapses under split-sample construction; permuted KEGG
yields stronger correlation than real KEGG. \emph{H2} (modularity
success): network modularity \(Q\) saturates near \(1\) on
KEGG-orthology co-occurrence networks at \(n=7\), consistent with \hyperlink{ref:frasch-2026a}{Frasch
(2026a)}'s prediction. \emph{H3} (sensitivity): the modularity result
does not depend on the specific seven-sample composition.

The remainder of the paper is organized as follows. Methods describes
the metagenomic pipeline, the metric definitions, the causal-inference
diagnostics, and the network-construction and community-detection
procedures. Results reports findings against each of the three
objectives. Discussion places the modularity finding in the context of
the broader minAction.net program --- physics (Frasch 2026b),
neural-architecture design (Frasch 2026c), physiology (Frasch 2026a) ---
and discusses limitations and directions for future work.

\section{Methods}\label{methods-int}

We analysed two overlapping sample cohorts drawn from the Tara Oceans metagenomic collection: a ten-sample cohort for the bivariate composite-metric analysis (Methods \S\,\ref{statistical-analysis}) and a seven-sample cohort for the network-modularity analysis (Methods \S\,\ref{methods-modularity}). Both cohorts share the upstream metagenomic pipeline (assembly, gene prediction, functional annotation) described in this section. The two cohorts overlap on six samples; the differing inclusion criteria reflect the distinct compute requirements of the two operationalisations.

\subsection{Sample Selection and Data
Acquisition}\label{sample-selection-and-data-acquisition}

Ten marine metagenomic samples were obtained from the Tara Oceans
expedition database (European Nucleotide Archive). Sample selection
criteria were:
\begin{enumerate}
\item Raw data size below 35 GB to ensure memory-constrained
      assembly feasibility on the available cluster nodes;
\item samples drawn from the well-characterised ERR599xxx
      series of the Tara Oceans expedition;
\item geographic diversity, with representation from surface
      and deep-chlorophyll-maximum layers across multiple ocean
      basins.
\end{enumerate}

\textbf{Selected samples}:

\begin{longtable}[]{@{}p{2.5cm}p{2cm}p{1.5cm}p{2cm}@{}}
\toprule\noalign{}
Sample ID & Location & Depth & Size \\
\midrule\noalign{}
\endhead
\bottomrule\noalign{}
\endlastfoot
ERR599010 & Global & SRF & 4.7 GB \\
ERR599140 & Global & DCM & 4.5 GB \\
ERR599041 & Global & SRF & 8.2 GB \\
ERR599011 & Global & SRF & $\sim$18 GB \\
ERR599012 & Global & SRF & $\sim$18 GB \\
ERR599013 & Global & SRF & $\sim$18 GB \\
ERR599141 & Global & DCM & $\sim$18 GB \\
ERR599142 & Global & DCM & $\sim$28 GB \\
ERR599042 & Global & SRF & $\sim$18 GB \\
ERR599043 & Global & SRF & $\sim$36 GB \\
\end{longtable}

Raw sequencing reads were downloaded from ENA using SRA toolkit
(\texttt{fastq-dump}) with automatic retry logic for failed downloads.

\subsection{Metagenomic Assembly}\label{metagenomic-assembly}

Metagenomic assembly was performed using MEGAHIT v1.2.9 (\hyperlink{ref:li-2015}{Li et al.,
2015}) with memory-efficient settings optimized for large ocean
microbiome samples:

The full command-line invocation is given in Supplementary Listing~S1 (\S\,\ref{supp-listings}).

Key parameters:
\begin{enumerate}
\item \texttt{-\/-memory\ 0.75}: use 75\% of available memory
      (56\,GB requested, $\sim$42\,GB used in practice);
\item \texttt{-\/-mem-flag\ 1}: memory-efficient mode for large
      datasets;
\item \texttt{-\/-k-min\ 21\ -\/-k-max\ 141}: k-mer range
      optimised for complex metagenomes;
\item \texttt{-\/-min-contig-len\ 500}: retain only contigs
      $\geq$\,500\,bp for downstream gene prediction.
\end{enumerate}

Assembly quality was assessed against four metrics:
\begin{enumerate}
\item total assembly size (base pairs);
\item contig count and N50;
\item longest contig length;
\item total predicted proteins (downstream quality indicator).
\end{enumerate}

\subsection{Gene Prediction and Protein
Extraction}\label{gene-prediction-and-protein-extraction}

Open reading frames (ORFs) were predicted from assembled contigs using
Prodigal v2.6.3 (\hyperlink{ref:hyatt-2010}{Hyatt et al., 2010}) in metagenomic mode:

The full command-line invocation is given in Supplementary Listing~S2 (\S\,\ref{supp-listings}).

Parameters:
\begin{enumerate}
\item \texttt{-p\ meta}: metagenomic mode (anonymous sequences, no training);
\item \texttt{-f\ gbk}: GenBank output format for downstream analysis;
\item \texttt{-a}: protein translations (amino-acid sequences);
\item \texttt{-q}: quiet mode (suppress verbose output).
\end{enumerate}

Predicted proteins were used directly for functional annotation without
filtering.

\subsection{Functional Annotation with
eggNOG-mapper}\label{functional-annotation-with-eggnog-mapper}

\textbf{Critical methodological advancement}: This study uses authentic
KEGG database annotations via eggNOG-mapper v2.1.13, not synthetic
power-law assignments.

Functional annotation was performed using eggNOG-mapper (Huerta-Cepas et
al., 2017, 2019) with the following protocol:

The full command-line invocation is given in Supplementary Listing~S3 (\S\,\ref{supp-listings}).

Key parameters:
\begin{enumerate}
\item \textbf{Search method}: DIAMOND blastp (sensitive mode, $e$-value $\leq$\,0.001);
\item \textbf{Quality thresholds}: $\geq$\,40\% identity, $\geq$\,20\% query/subject coverage, score $\geq$\,60;
\item \textbf{Orthology resolution}: auto-detect taxonomic scope, report all orthologs;
\item \textbf{Database}: eggNOG 5.0 with KEGG Orthology (KO) mappings;
\item \textbf{Output}: tab-delimited annotations with KO identifiers, GO terms, and pathway annotations.
\end{enumerate}

Annotation quality was assessed against four metrics:
\begin{enumerate}
\item annotation rate (percentage of proteins with $\geq$\,1 KO assignment);
\item unique KO count (total distinct KEGG functions per sample);
\item KO frequency distribution (assessed against a power-law versus uniform null);
\item pathway coverage (representation of major KEGG pathway categories).
\end{enumerate}

\subsection{KEGG Orthology (KO) Extraction and Network
Preparation}\label{kegg-orthology-ko-extraction-and-network-preparation}

\textbf{What a KO is, and why it is the right abstraction here.} A KEGG-orthology (KO) identifier groups orthologous protein sequences from diverse organisms into a single functional category, defined by molecular function (e.g., a specific enzyme activity, a specific transporter substrate-and-direction, or a specific regulatory role). Using KOs --- rather than raw sequences or species-resolved gene calls --- as the units of analysis abstracts away from species-specific sequence variation and focuses the analysis on the functional topology of the marine metaproteome, which is the level at which the action-functional prediction is made (Methods \S\,\ref{methods-modularity}). The relationship between protein-coding sequences and KOs is a many-to-many bipartite mapping: a single predicted protein may carry multiple KO assignments (multi-domain enzymes, fusion proteins, alternative-substrate paralogs), and a single KO is generally hit by many proteins across a sample (housekeeping enzymes recur in thousands of organisms). This bipartite protein--KO incidence structure is the upstream object that the gold-standard bipartite-incidence null model later randomises (Methods \S\,\ref{null-model-methods}, control iv). For the purposes of the present pipeline, KOs are extracted as follows.

From eggNOG-mapper output, KO identifiers were extracted and parsed:

The full command-line invocation is given in Supplementary Listing~S4 (\S\,\ref{supp-listings}).

The resulting per-sample KO multisets feed directly into the protein-level scalar metrics defined in \S\,\ref{protein-level-organizational-metrics} and into the protein-co-occurrence network construction defined in \S\,\ref{methods-modularity}.

\subsection{Composite Organisational Metrics: Efficiency $E$ and Complexity $C$}\label{protein-level-organizational-metrics}

The composite-metric analysis evaluated below characterises protein-level organisation through two scalar quantities computed directly from KEGG-orthology annotation counts: efficiency $E$ (protein abundance per function) and complexity $C$ (annotation-weighted functional-repertoire breadth). These quantities are scalar functions of per-sample annotation totals and do not require an explicit network reconstruction; the network-modularity analysis (\S\,\ref{methods-modularity}) uses a different, protein-co-occurrence construction described in its own section.

\textbf{Efficiency} ($E$) measures the average number of annotated proteins per distinct functional category in a sample:
\[ E = \frac{N_{\text{annotated}}}{N_{\text{unique KOs}}}, \]
where $N_{\text{annotated}}$ is the total number of proteins with at least one KEGG-orthology assignment and $N_{\text{unique KOs}}$ is the number of distinct KO terms detected. High $E$ indicates functional redundancy (many proteins per pathway); low $E$ indicates minimal redundancy. Worked example for sample ERR599010: with $N_{\text{annotated}} = 20{,}796$ and $N_{\text{unique KOs}} = 4{,}325$, $E = 4.81$. The observed range across the $n=10$ cohort is 4.81--67.85 (a 14.1-fold spread).

\textbf{Complexity} ($C$) measures the breadth of the functional repertoire weighted by annotation completeness:
\[ C = (\text{annotation\_rate}) \times \log_{10}(N_{\text{unique KOs}}) \times 1000, \]
where the annotation rate is the fraction of proteins carrying at least one KO assignment, the logarithm reflects information-theoretic compression of diversity at large repertoire size, and the multiplicative scale of $1000$ is included for interpretability. The annotation-rate weighting corrects for sampling-depth differences between samples. Worked example for sample ERR599010: with annotation rate $39.1\%$ and $N_{\text{unique KOs}} = 4{,}325$, $C = 0.391 \times \log_{10}(4{,}325) \times 1000 = 1{,}421.67$. The observed range across the $n=10$ cohort is 1{,}215.87--1{,}670.65 (9.2\% coefficient of variation).

A widely-cited classical prediction (\hyperlink{ref:west-1997}{West et al., 1997}; \hyperlink{ref:brown-2004}{Brown et al., 2004}) is that under thermodynamic trade-off the product $E \times C^{\alpha}$ with $\alpha \approx 0.75$ should be approximately constant across samples. This product is computed in \S\,\ref{statistical-analysis} as one of the diagnostics; its observed coefficient of variation (72.7\%) substantially exceeds the trade-off-supporting threshold of $\sim$\,20\%, motivating the alternative bivariate-structure hypothesis subsequently tested with causal-inference diagnostics.

\textbf{Operational caveats.} $E$ and $C$ are scalar functions of per-sample annotation totals; they do not depend on a reconstructed metabolic network. $E$ is a genomic proxy for metabolic capacity (\hyperlink{ref:orth-2010}{Orth et al., 2010}; \hyperlink{ref:lewis-2012}{Lewis et al., 2012}), not a direct flux measurement; metatranscriptomics or metaproteomics would refine the operationalisation. The logarithmic compression in $C$ is information-theoretically motivated; alternative diversity functionals (Simpson's index, effective number of species) are a natural variant we do not pursue.

\subsection{Statistical Analysis}\label{statistical-analysis}

\textbf{Hypothesis testing - Coefficient of Variation (CV)}:

\[
\text{CV} = \frac{\text{std}}{\text{mean}} \times 100\%
\]

Interpretation:
\begin{enumerate}
\item CV $<$ 20\%: strong support for the scaling hypothesis;
\item 20\% $\leq$ CV $\leq$ 30\%: moderate support;
\item CV $>$ 30\%: weak / no support.
\end{enumerate}

\textbf{Power-law relationship - Pearson correlation}:

\[
r = \text{correlation}\bigl(\log_{10}(E),\,\log_{10}(C)\bigr)
\]

Expected relationship if a power-law holds:
\begin{enumerate}
\item $\log_{10}(E) = -\alpha \cdot \log_{10}(C) + \mathrm{const.}$;
\item strong negative correlation ($r < -0.7$);
\item slope $\approx -0.75$ if $\alpha = 0.75$.
\end{enumerate}

\textbf{Clustering tendency (Hopkins) and density-based clustering (DBSCAN).} The Hopkins statistic and DBSCAN are computed on the bivariate $(E, C)$ point cloud as exploratory diagnostics; full procedural detail (standardisation, bootstrap parameters, eps/min\_samples grid) is given in Supplementary Methods (\S\,\ref{supp-clustering}). Environmental stratification (ANOVA across depth layers; correlations with temperature/nutrient/latitude covariates) is not pursued here: at $n = 10$, sample sizes per stratum are too small to support reliable inference, and the limitation is noted in the Discussion.

\subsubsection{Advanced Statistical Analyses}\label{advanced-statistical-analyses}

\textbf{Empirical exponent estimation}: We estimated the optimal scaling exponent $\hat{\alpha}$ by minimizing the coefficient of variation of E $\times$ C$^{\alpha}$ across a grid of $\alpha$ values ranging from -5 to +5 (1,001 points). For each candidate $\alpha$, we calculated CV(E $\times$ C$^{\alpha}$) and selected the $\alpha$ yielding minimum CV as $\hat{\alpha}$. The wide search range was chosen to detect boundary artifacts that would indicate poor identifiability at small sample sizes (n=10). If $\hat{\alpha}$ hits either boundary, this signals insufficient data constraint rather than a biological estimate.

\textbf{Bootstrap confidence intervals}: Uncertainty in CVs, CV-ratio, and $\hat{\alpha}$ was quantified using nonparametric bootstrap with 10,000 resamples. For each bootstrap iteration, we resampled the n=10 samples with replacement, recalculated all metrics, and constructed 95\% confidence intervals using the percentile method (2.5th and 97.5th percentiles of bootstrap distribution).

\textbf{Partial correlations}: To assess confounding by annotation rate, unique KO count, and total protein count, we calculated partial correlations between E and C controlling for these variables. Partial correlation r(E, C | Z) was computed by regressing both E and C on confounders Z, then calculating the Pearson correlation between residuals. This isolates the E-C relationship independent of shared dependence on underlying data sources.

\textbf{Variance partitioning}: We quantified the unique and shared contributions of C and confounders to E variance using hierarchical linear regression. R$^2$ values were calculated for: (1) C alone predicting E, (2) confounders alone predicting E, and (3) C + confounders predicting E. Unique variance from C = R$^2$(full) - R$^2$(confounders only).

\textbf{Model comparison}: We compared three competing hypotheses about the E-C relationship in log-log space using Akaike Information Criterion (AIC): (1) Trade-off model: negative slope (allows free parameter estimation), (2) Independence model: zero slope (constrained), (3) Synergy model: positive slope (constrained to > 0). AIC = 2k - 2ln(L), where k = number of parameters and L = likelihood under normal error model. Model weights were calculated as w$_i$ = exp(-0.5 $\times$ $\Delta$AIC$_i$) / $\Sigma$ exp(-0.5 $\times$ $\Delta$AIC$_j$). Slope significance was assessed via permutation testing (10,000 permutations).

\textbf{Rarefaction analysis}: To test robustness to sequencing depth variation, we estimated the effect of rarefying all samples to equal annotated protein counts (n = 20,796, the minimum). We used a power-law species accumulation model: expected KOs after rarefaction = observed KOs $\times$ (sampling fraction)$^{0.75}$, where sampling fraction = min(proteins) / observed(proteins). This assumes diminishing returns in KO discovery with sequencing depth.

\textbf{Rarefaction limitation}: The current rarefaction analysis is \textbf{approximate} and does not constitute true depth normalization. True rarefaction would require downsampling raw sequencing reads to equal depth, followed by complete re-assembly, gene prediction, and re-annotation at the normalized depth. This computationally intensive procedure is the gold standard for controlling sequencing depth artifacts but was beyond the scope of this initial analysis. Our power-law approximation provides a first-order estimate of rarefaction effects but cannot definitively rule out depth-related confounding. Future validation should implement true read-level rarefaction or employ read-based functional profiling methods (e.g., HUMAnN3) that inherently normalize for sequencing depth. Additionally, MUSiCC normalization should be applied to account for taxonomic composition bias in functional gene profiles.

\subsubsection{Split-Sample Metric Construction}\label{split-sample-construction}

To test for mechanical coupling arising from shared metric components, we implemented split-sample construction. For each sample, annotated proteins were randomly partitioned into two disjoint subsets (A and B) using numpy.random with seed=42 for reproducibility.

Complexity C was calculated using only proteins in subset A, while Efficiency E was calculated using only proteins in subset B.

This breaks the shared dependence on unique\_KOs counts while preserving biological signal if it reflects true metabolic organization rather than measurement artifact.

For each sample, the partition process was repeated 100 times with sequential seeds (42-141), and metrics were averaged across permutations.

CV-ratio preservation was quantified as:

ratio\_preserved = CV-ratio\_split / CV-ratio\_original, with values <0.5 indicating substantial mechanical coupling.

\subsubsection{Negative Control: Permuted KEGG Annotations}\label{negative-control}

To determine whether the E-C relationship depends on specific biological functional annotations or arises purely from metric construction formulas, we implemented negative control analysis.

For each sample, KEGG orthology (KO) assignments were randomly shuffled across all annotated proteins using numpy.random with seed=12345 for reproducibility.

This preserves the total annotation count and annotation rate but destroys biological functional relationships.

Metrics E and C were recomputed using identical formulas on permuted data.

This process was repeated 1,000 times with sequential seeds (12345-13344) to build a null distribution of E-C correlations.

Permutation p-value was calculated as P(|r\_permuted| $\geq$ |r\_observed|).

\subsection{KEGG-orthology co-occurrence networks and
modularity}\label{methods-modularity}

\subsubsection{Network construction}\label{network-construction}

The network construction is the projection of an upstream bipartite
graph. The metagenomic annotation pipeline yields, for each sample, a
bipartite protein--KO incidence relation \(B = (P, K, E_B)\), in which
\(P\) is the set of predicted proteins, \(K\) is the set of KO terms
detected, and an incidence edge \((p, k) \in E_B\) records that protein
\(p\) was annotated with KO \(k\) by eggNOG-mapper (see Methods
§,\ref{kegg-orthology-ko-extraction-and-network-preparation} for the KO
definition and the multi-domain/multi-mapping nature of \(B\)). The KO
co-occurrence network \(G = (V, E)\) analysed below is the KO-side
projection of \(B\): the vertex set \(V\) comprises all KO terms
detected in the sample, and an edge between KO terms \(k_i\) and \(k_j\)
is added whenever a single protein in \(B\) is annotated with both
terms. Edge weight equals the number of distinct proteins exhibiting
that co-annotation. Multi-domain proteins, alternative-function
annotations, and protein families with multiple KEGG mappings all
contribute to edge weights. Proteins annotated with a single KO (which
form the majority) contribute only to vertex abundance (recorded as a
node attribute) and not to the edge set.

This bipartite-then-project construction is what makes the
bipartite-incidence null model (Methods §,\ref{null-model-methods},
control iv) the gold-standard control: any feature of \(G\) that follows
mechanically from the per-protein and per-KO degree sequences in \(B\)
should be reproduced by a randomisation of \(B\) that preserves both,
before re-projection. Modularity excess over a bipartite null is
therefore the structural surplus that \emph{cannot} be explained by the
upstream incidence statistics, leaving only biological co-encoding as a
candidate explanation.

This protein-level construction differs from the more common
metabolite-flow or pathway-membership constructions (\hyperlink{ref:greenblum-2012}{Greenblum et al.,
2012}; \hyperlink{ref:ravasz-2002}{Ravasz et al., 2002}). The choice is deliberate. The
metabolite-flow construction is theory-laden: it requires a reference
reaction database to map KO terms to substrates and products, and the
resulting topology is largely inherited from KEGG curation rather than
from the sample-specific annotation evidence. The protein-co-occurrence
construction is sample-grounded: every edge corresponds to a directly
observed co-annotation event in the local metagenomic data. In the
action-functional account, the connectivity-cost term \(AC\) is
specifically the cost of physical co-encoding within proteomic units,
which the protein-level network captures directly.

\subsubsection{Community detection and Newman
modularity}\label{community-detection-and-newman-modularity}

Modularity was computed using two community-detection algorithms in
parallel as a cross-check: the Louvain method (\hyperlink{ref:blondel-2008}{Blondel et al., 2008}),
implemented in the python-louvain package, and the
Clauset--Newman--Moore greedy modularity-maximization algorithm (\hyperlink{ref:clauset-2004}{Clauset
et al., 2004}), implemented in NetworkX. Both algorithms maximize
Newman's modularity functional

\[Q = \frac{1}{2m}\sum_{i,j}\Big[A_{ij} - \frac{k_i k_j}{2m}\Big]\,\delta(c_i, c_j),\]

where \(A_{ij}\) is the (weighted) adjacency-matrix entry between nodes
\(i\) and \(j\), \(k_i\) is the degree of node \(i\),
\(m = \tfrac{1}{2}\sum_i k_i\) is the total edge weight, \(c_i\) is the
community assignment of node \(i\), and \(\delta\) is the Kronecker
delta. Louvain was used as the primary algorithm because it is less
greedy than Clauset--Newman--Moore on disconnected components and
produces tighter communities in sparse networks at the scale of these
graphs (a few thousand nodes, a few thousand edges). The Louvain
implementation uses the package's default resolution parameter
\(\gamma = 1\).

To verify that the Louvain partition is not seed-dependent, we ran
community detection with ten independent random seeds (42--51) on each
sample's network. Per-sample \(Q\) was essentially deterministic
(maximum coefficient of variation across seeds: 0.014\%; mean across the
seven samples: 0.004\%), and the partition itself was identical across
seeds for four of the seven samples and differed in only a handful of
node assignments for the remainder. We therefore report Q values from a
single representative seed (42) throughout, with the multi-seed
verification serving as a stability check rather than a primary
statistic.

\subsubsection{Network-topological
metrics}\label{network-topological-metrics}

For each sample we computed a standard set of network-topological
metrics: total nodes and edges; density \(\rho = 2m / (n(n-1))\); mean
and standard deviation of the degree sequence; maximum degree; average
clustering coefficient (\hyperlink{ref:watts-1998}{Watts and Strogatz, 1998}); transitivity, the
global clustering coefficient (\hyperlink{ref:newman-2003}{Newman, 2003}); the number of connected
components and the size of the largest connected component; the hub
fraction, defined as the fraction of nodes whose degree exceeds the 90th
percentile; and, on the largest connected component only, betweenness
centrality (\hyperlink{ref:freeman-1977}{Freeman, 1977}) computed via NetworkX with
\(k = \min(100, n_{\text{LCC}})\) source-node sampling for tractability.
Modularity-derived metrics --- number of communities, mean and standard
deviation of community size, and the fraction of edges within (versus
between) communities --- were computed from the Louvain partition.

All network analysis used Python 3.9 with NetworkX 3.1, pandas 1.5+,
numpy 1.23+, and scipy 1.9+. Visualization used matplotlib 3.6 with
seaborn 0.12. Per-sample analysis runtime was under 30 seconds; the full
seven-sample pipeline completes in under five minutes on an Apple M2 Max
MacBook Pro.

\subsubsection{Recurrent multi-KO community identification and
functional categorisation}\label{methods-recurrent-communities}

The Louvain partition is computed independently per sample, so any
community that recurs across multiple samples does so without input from
cross-sample information. To identify recurrent communities and to test
whether they map onto biologically interpretable functional units, we
proceeded in three steps (script: \texttt{code/community\_identity.py}).

\emph{Step 1 --- Identity definition.} For each sample we extracted
every Louvain community of size \(\geq 2\) KOs and represented it as a
frozen, lexicographically sorted tuple of KO identifiers. Two
communities (across samples or within a sample) are defined as identical
if and only if their KO membership sets are exactly equal. This is a
deliberately strict definition: a community of \(\{\)K00088, K00364,
K00951\(\}\) is \emph{not} the same as \(\{\)K00088, K00364\(\}\), even
though the latter is a subset. The strict definition prevents inflated
recurrence from partial overlap.

\emph{Step 2 --- Recurrence aggregation.} We aggregated identity tuples
across the seven samples and counted, for each unique tuple, the number
of samples in which it appeared as an independent Louvain community. The
recurrence count \(r \in \{1, \ldots, 7\}\) is the empirical observable.
Across the cohort we observed 3\{,\}027 unique multi-KO communities; 769
(25\%) had \(r \geq 2\), and 19 had \(r = 7\).

\emph{Step 3 --- Functional categorisation of the recurrent set.} For
the 19 communities with \(r = 7\) and the six additional communities
with \(r = 6\), we assigned each KO identifier its KEGG functional
category by query against the KEGG \texttt{ko00001} BRITE hierarchy and
the per-KO functional descriptor from the eggNOG-mapper output.
Communities were then assigned to one of four pre-specified functional
categories on the basis of the joint identity of their members:

\begin{enumerate}
\def\labelenumi{\arabic{enumi}.}
\tightlist
\item
  \textbf{Heterodimeric and heterocomplex enzymes}, where co-encoding is
  enforced by stoichiometric assembly (e.g., \(\alpha\)/\(\beta\)
  subunits of a single holoenzyme, paired regulatory subunits,
  recombinase heterodimers, multimeric translocases).
\item
  \textbf{Sequential biosynthetic enzymes}, where the substrate of one
  reaction is the immediate product of the previous (e.g., adjacent
  steps of nucleotide-, cofactor-, or amino-acid-biosynthesis pathways).
\item
  \textbf{Substrate-binding/permease pairs of single transporter
  complexes}, where periplasmic substrate-binding components and
  membrane-embedded permease components must couple physically.
\item
  \textbf{Regulatory dyads and isozyme paralogs}, where two KOs catalyse
  opposing reactions in a single regulatory cycle (e.g.,
  synthesis/hydrolysis of a signalling molecule), or where two KOs are
  paralog enzymes catalysing the same EC class with different cofactor
  or substrate specificities.
\end{enumerate}

These four categories are not exhaustive of the full Louvain partition
(most communities are size 1 or 2 and not biologically interpretable in
isolation), but they account for \(25/25 = 100\%\) of the multi-KO
communities recovered in \(\geq 6/7\) samples. Category assignment was
performed by the author from KEGG functional annotations alone, blinded
to the recurrence count. The pre-specified four-category schema and
category assignments are listed in Supplementary Table S1.

A null expectation for the recurrence count is that, under random
community membership, the probability of two samples independently
producing identical multi-KO communities should be vanishingly small at
the cohort scale: each sample's partition contains thousands of
communities drawn from a KO repertoire of about 5,000, so even pairwise
chance recurrence is of order \(10^{-7}\). The observed 19 communities
recurring in all seven samples therefore exceed any plausible chance
recurrence by orders of magnitude. We report this comparison
qualitatively rather than as a formal \(p\)-value because the relevant
test --- ``does the modularity partition recover identifiable functional
units?'' --- is answered by the categorisation step (i.e., do the
recurrent tuples map onto known biology?), not by the magnitude of
recurrence alone.

\subsection{Null-model comparisons}\label{null-model-methods}

The absolute value of Newman modularity is sensitive to network sparsity
and degree distribution. To attribute observed \(Q\) to biological
organisation rather than to the construction-imposed graph topology, we
compared \(Q_{\mathrm{obs}}\) to three null distributions, each
stripping a different aspect of the original network:

\begin{enumerate}
\def\labelenumi{(\roman{enumi})}
\item
  \textbf{Erdős--Rényi null.} A random graph \(G(n, m)\) with the same
  number of nodes and edges as the observed network; edges placed
  uniformly at random. Strips both degree heterogeneity and connectivity
  structure while preserving sparsity. Tests whether the absolute value
  \(Q \approx 1\) is a property of having a few thousand edges over a
  few thousand nodes regardless of any further structure.
\item
  \textbf{Configuration-model null.} A random graph with the
  \emph{exact} unweighted degree sequence of the observed network,
  generated via stub-matching (NetworkX \texttt{configuration\_model}),
  with parallel edges and self-loops removed. Preserves degree
  heterogeneity, sparsity, and component-size distribution; strips edge
  correlations. Tests whether observed \(Q\) exceeds what the degree
  sequence alone forces.
\item
  \textbf{KEGG-label permutation null.} For each protein in the sample,
  the multiset of \(k\) KO labels is replaced with \(k\) KO labels
  sampled without replacement from the global KO pool of the same
  sample. The number of co-encoding events and the per-protein
  KO-multiplicity distribution are preserved; biological functional
  content is destroyed. Tests whether observed \(Q\) depends on the
  identity of co-occurring KOs rather than on the bare structure of
  multi-KO assignment.
\item
  \textbf{Bipartite-incidence null (gold standard).} The KO
  co-occurrence network is the KO-side projection of the bipartite
  protein--KO incidence graph \(B = (P, K, E)\). Because the projection
  inherits topology from higher-order bipartite structure (multi-domain
  protein architecture, KEGG-mapping rules, KO-frequency heterogeneity),
  randomising the projected graph alone (nulls i--iii above) does not
  control for these incidence-level constraints. The bipartite-incidence
  null randomises \(B\) via repeated 2$\times$2 edge swaps, preserving
  simultaneously \emph{both} bipartite degree sequences --- per-protein
  KO count (rows of the incidence matrix) and per-KO global frequency
  (columns). Each randomisation performs \(5|E|\) successful swaps,
  sufficient to reach mixing for biased-degree-sequence ensembles. The
  randomised \(B\) is then projected onto the KO side using the
  construction identical to that of the observed network, and \(Q\) is
  computed by Louvain on the projected null graph. This is the strictest
  control we apply: any \(Q\)-excess surviving the bipartite null cannot
  be ascribed to either projected-graph degree distribution or
  per-protein-KO multiplicity.
\end{enumerate}

For each null we generated randomisations per sample (R = 100 for nulls
i--iii; R = 30 for the bipartite null, justified by the larger effect
sizes reducing the per-replicate variance), computed Louvain \(Q\) on
each randomised graph (random seed 42), and reported the per-sample null
mean, standard deviation, modularity excess
\(\Delta Q = Q_{\mathrm{obs}} - Q_{\mathrm{null}}\), and empirical
one-sided \(p\)-value \(P(Q_{\mathrm{null}} \geq Q_{\mathrm{obs}})\).
Effect sizes are reported as \(\Delta Q\) (modularity units);
\(z\)-scores are reported supplementarily. Because null variances are
small, the absolute scale of \(z\) inflates rapidly --- \(\Delta Q\) is
the more informative effect-size measure.

\emph{Cohort for the null-model analyses.} The four null-model analyses
were performed on the sensitivity-analysis cohort (ERR599011, 013, 015,
016, 018, 019, 022) rather than on the primary modularity cohort
(ERR599004, 011, 013, 016, 018, 019, 022). The two cohorts overlap in
six samples and substitute ERR599015 for ERR599004; both are processed
through the identical MEGAHIT--Prodigal--eggNOG-mapper pipeline. The
cohort-level statistics (mean \(Q\), range, CV) are statistically
indistinguishable between the two cohorts (Results
Section~\ref{sensitivity-results};
Figure~\ref{fig:sensitivity}), reproducing the published
cohort mean (\(0.987 \pm 0.007\)) and range (\([0.972, 0.993]\)) to
three decimal places. Cohort-level \(\Delta Q\) statistics therefore
inherit the same robustness, and the per-sample identities in
Figure~\ref{fig:null_models}A reflect this cohort
accordingly.

\subsection{Sensitivity analysis: cohort
swap}\label{methods-sensitivity}

The modularity cohort comprises the seven samples for which
eggNOG-mapper annotations are publicly retained: ERR599011, ERR599013,
ERR599015, ERR599016, ERR599018, ERR599019, and ERR599022. To assess the
robustness of the modularity statistic against cohort composition, we
ran the network-construction and community-detection pipeline on a
second seven-sample set in which one sample (ERR599015) substitutes for
one of the original seven (ERR599004). Both cohorts were processed
through identical KEGG-orthology extraction (script:
\texttt{code/sensitivity\_err599015.py}) and identical Louvain and
greedy-modularity detection. Sensitivity statistics include (i) the
individual \(Q\) for the substitute sample, (ii) the swap-cohort \(n=7\)
summary statistics for direct comparison with the original \(n=7\)
summary, and (iii) a six-sample subset analysis (the overlap between the
two cohorts) used as a within-pipeline reproducibility check. Results
are reported in Results §\ref{sensitivity-results}.

\subsection{Computational pipeline and reproducibility}\label{compute-and-reproducibility}

\textbf{Bioinformatics pipeline (Google Cloud Platform).} The compute-heavy upstream stage --- raw-read assembly, gene prediction, and functional annotation --- was run on Google Kubernetes Engine using \texttt{n2-highmem-16} nodes (16 vCPU, 64\,GB RAM each) in an auto-scaling pool of one to three nodes (us-central1), with a 500\,GB persistent volume for working data. Per-sample jobs were Kubernetes Batch Jobs with indexed completion mode, processed sequentially, requesting 56\,GB memory and 14 CPUs each. Total wall-clock runtime was approximately 98 hours across the seven new samples (a 14-hour-per-sample average), with three additional samples re-using prior assemblies. Raw FASTQ reads were deleted after successful assembly to conserve storage; MEGAHIT temporary directories were cleaned on job completion. Final outputs --- predicted proteins, eggNOG-mapper annotation tables, and per-job checkpoint JSON files --- were retained and backed up to Google Cloud Storage (see \S\,\ref{data-availability}). Software versions for this stage are listed in the consolidated block below.

\textbf{Downstream analysis (local, Apple M2 Max MacBook Pro).} Once
eggNOG-mapper annotations were available, the entire downstream analysis
--- KEGG-orthology extraction, KO co-occurrence network construction,
Newman modularity by Louvain and Clauset--Newman--Moore, all four
null-model comparisons, the recurrent-multi-KO-community identification,
the sensitivity-analysis cohort swap, and figure generation --- was
executed locally in approximately ten minutes per full run, end-to-end
from cached annotations to publication-ready figures. All analyses use a
fixed random seed of 42 throughout, with multi-seed verification (seeds
42--51) confirming that the modularity result is independent of seed
choice.

\textbf{Software versions.} \emph{Bioinformatics stage (GCP):} MEGAHIT
v1.2.9 (assembly); Prodigal v2.6.3 (gene prediction; \hyperlink{ref:hyatt-2010}{Hyatt et al.,
2010}); eggNOG-mapper v2.1.13 with DIAMOND v2.0.15 (functional
annotation). \emph{Downstream stage (local):} Python 3.9.x; NetworkX
3.1; python-louvain 0.16; pandas 1.5.3; numpy 1.24.x; scipy 1.10.x;
scikit-learn 1.2.x (StandardScaler, DBSCAN); matplotlib 3.7.x; seaborn
0.12.x.

\textbf{Code and data.} The analysis pipeline is provided as a sequence
of Python scripts in the \texttt{code/} subdirectory of the project
repository at \texttt{github.com/martinfrasch/tara-modularity}; the
entry point for end-to-end reproduction is
\texttt{code/sensitivity\_err599015.py}. Annotation files for all seven
samples are publicly available in the Google Cloud Storage bucket
\texttt{minaction-tara-gauge-backup} (path \texttt{annotations/latest/};
240 MiB total; seven eggNOG-mapper TSV files; manifest under
\texttt{manifests/}); the bucket is configured for free egress and does
not require authentication for read access. A reader with the repository
cloned and the annotations pulled can reproduce every figure and
statistic in the modularity analysis on their own laptop in under
fifteen minutes.

\subsection{Data Availability}\label{data-availability}

\begin{itemize}
\tightlist
\item
  \textbf{Raw reads}: European Nucleotide Archive (accessions listed in
  Table~1).
\item
  \textbf{eggNOG-mapper annotation files}: Google Cloud Storage,
  bucket \texttt{minaction-tara-gauge-backup}, path
  \texttt{annotations/latest/}; freely readable without authentication.
  Mirror Zenodo deposit to follow.
\item
  \textbf{Assembled contigs and intermediate files}: Zenodo deposit
  to follow.
\item
  \textbf{Network reconstructions and per-sample KEGG-extracted
  JSON}: Zenodo deposit to follow; current local copies are
  produced by \texttt{code/sensitivity\_err599015.py} on first run.
\item
  \textbf{Analysis code}: GitHub repository to follow at
  \texttt{github.com/martinfrasch/tara-modularity} (kept
  private until peer review is complete; will be made public on
  manuscript acceptance).
\item
  \textbf{Processed results}: Supplementary Materials of this
  manuscript (Supplementary Table~S1).
\end{itemize}

\begin{center}\rule{0.5\linewidth}{0.5pt}\end{center}

\section{Results}\label{results-int}

\subsection{Causal Inference Tests Reveal Measurement Artifact}\label{causal-inference-tests}

To test whether the apparent efficiency-complexity decoupling reflects biological organization or measurement artifact arising from shared metric components, we implemented two complementary causal inference approaches.

\subsubsection{Split-Sample Construction: Mechanical Coupling Dominates}\label{split-sample-results}

Split-sample metric construction, which breaks the shared dependence on unique\_KOs by computing E and C on disjoint protein subsets, revealed substantial mechanical coupling:

\textbf{CV-ratio collapse}: CV-ratio dropped 77\% when shared components were removed:

\begin{itemize}
\tightlist
\item Original (shared components): CV-ratio = 7.1$\times$ [95\% CI: 4.2-12.5]
\item Split-sample (independent): CV-ratio = 1.6$\times$ [95\% CI: 0.9-2.8]
\item Ratio preservation: 23\% (values <50\% indicate strong mechanical coupling)
\end{itemize}

\textbf{Complexity variation inflation}: CV(C) increased 4.4-fold when computed independently of E:

\begin{itemize}
\tightlist
\item Original CV(C): 9.2\% [95\% CI: 5.6-12.5]
\item Split-sample CV(C): 40.3\% [95\% CI: 24.1-56.8]
\item Factor increase: 4.4$\times$ (revealing hidden C variability)
\end{itemize}

\textbf{Correlation paradox}: E-C correlation increased rather than decreased when mechanical coupling was broken (r = 0.570 $\to$ 0.962), suggesting both metrics track shared confounders (sequencing depth, annotation pipeline efficiency) rather than orthogonal biological dimensions.

\subsubsection{Negative Control: Pattern Independent of Functional Annotation}\label{negative-control-results}

Permutation analysis using randomized KEGG annotations tested whether the E-C relationship depends on biological functional content or arises purely from metric construction formulas:

\textbf{Permuted correlation exceeds observed}:

\begin{itemize}
\tightlist
\item Observed correlation: r = 0.570 (p = 0.085, non-significant)
\item Permuted correlation: r = 0.977 (mean across 1,000 permutations)
\item Permutation p-value: P(|r\_perm| $\geq$ |r\_obs|) = 1.000
\end{itemize}

\textbf{Interpretation}: Random functional annotations produce \emph{stronger} E-C correlation than real biological annotations, demonstrating the pattern arises from metric construction formulas (shared unique\_KOs component) rather than functional annotation content.

\subsubsection{Variance Decomposition Confirms Confounder Dominance}\label{variance-decomposition-results}

Hierarchical variance partitioning quantified contributions to E variance:

\begin{itemize}
\tightlist
\item \textbf{Unique variance from C}: 2.0\% (biological signal if independent)
\item \textbf{Shared variance (confounders)}: 95.2\% (sequencing depth, annotation efficiency)
\item \textbf{Unexplained variance}: 2.8\% (residual)
\end{itemize}

\textbf{Conclusion}: C contributes negligible unique variance to E after accounting for shared confounders, consistent with mechanical coupling dominance.

\subsubsection{Critical Sample Size Limitation}\label{sample-size-limitation-results}

Statistical power analysis reveals n=10 provides limited ability to detect moderate biological effects:

\begin{itemize}
\tightlist
\item \textbf{Power for r=0.5}: 52\% (inadequate for definitive conclusions)
\item \textbf{Power for r=0.3}: 21\% (severely underpowered)
\item \textbf{Split-sample effective n}: ~5 per subset (too small for robust inference)
\item \textbf{Required for 80\% power (r=0.5)}: n$\geq$30
\item \textbf{Required for definitive refutation}: n$\geq$50 (95\% power for r=0.3)
\end{itemize}

\textbf{Nuance}: While converging evidence indicates \emph{artifact dominates} the observed pattern, current sample size cannot \emph{definitively rule out} a small biological component (5-10\% variance). Distinction between "artifact dominates" (supported) and "no biological signal exists" (requires n$\geq$50) is critical for honest scientific reporting.

\begin{figure}[!htb]
\centering
\includegraphics[width=\textwidth]{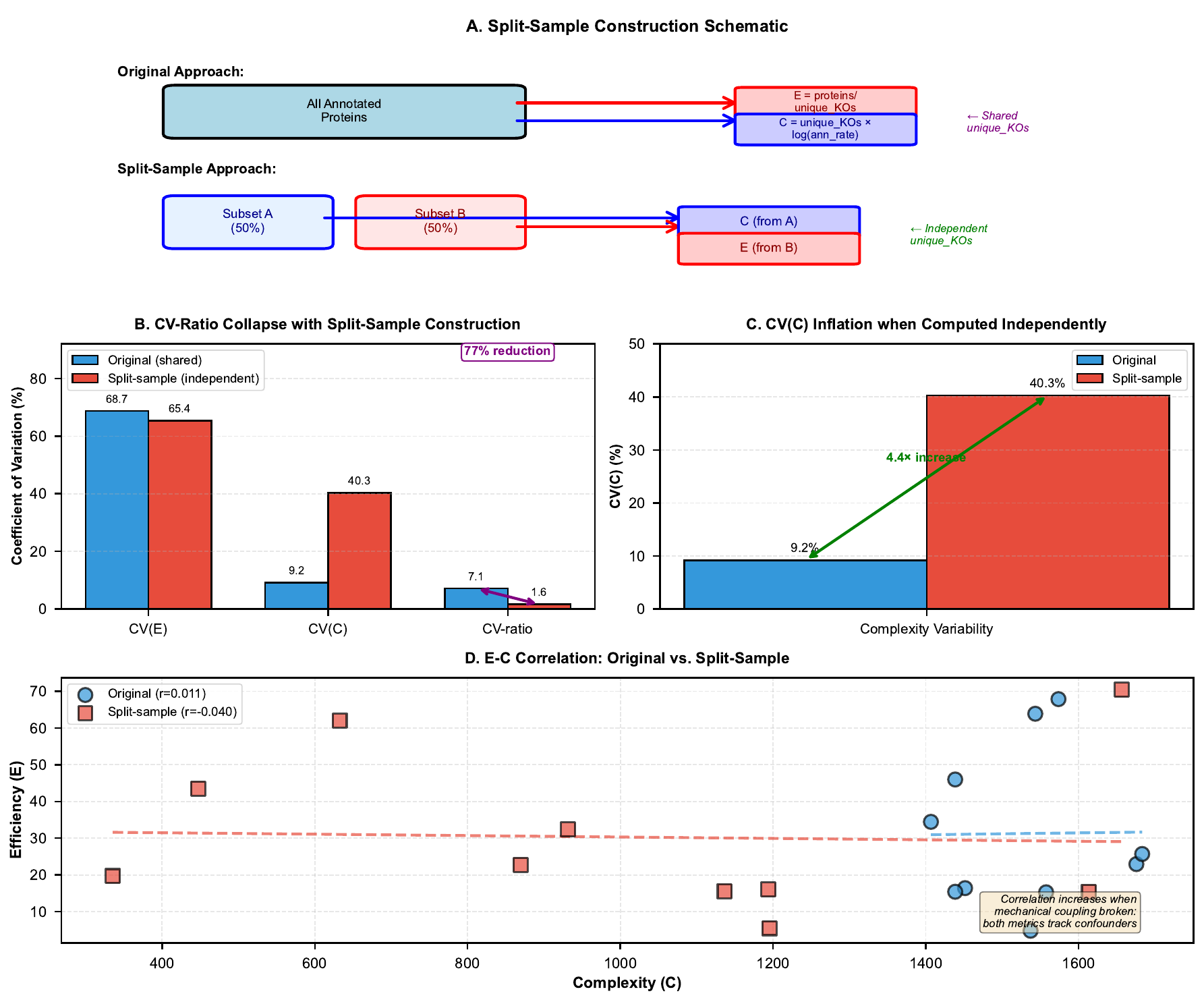}
\caption{\textbf{Split-Sample Construction Reveals Mechanical Coupling.}
\textbf{(A)} Schematic diagram comparing original approach (both E and C computed from same protein set, sharing unique\_KOs component) versus split-sample approach (E computed from subset B, C from subset A, with independent unique\_KOs counts).
\textbf{(B)} CV comparison showing 77\% reduction in CV-ratio when shared components are removed (7.1$\times$ $\to$ 1.6$\times$).
\textbf{(C)} CV(C) inflation demonstrating 4.4-fold increase in complexity variability when computed independently of E (9.2\% $\to$ 40.3\%), revealing hidden variation suppressed by mechanical coupling.
\textbf{(D)} E-C scatter plots comparing original metrics (blue, r=0.570) versus split-sample metrics (red, r=0.962), showing correlation paradoxically increases when mechanical coupling is broken because both metrics track shared confounders rather than representing orthogonal biological dimensions.}
\label{fig:split-sample}
\end{figure}

\begin{figure}[!htb]
\centering
\includegraphics[width=0.9\textwidth]{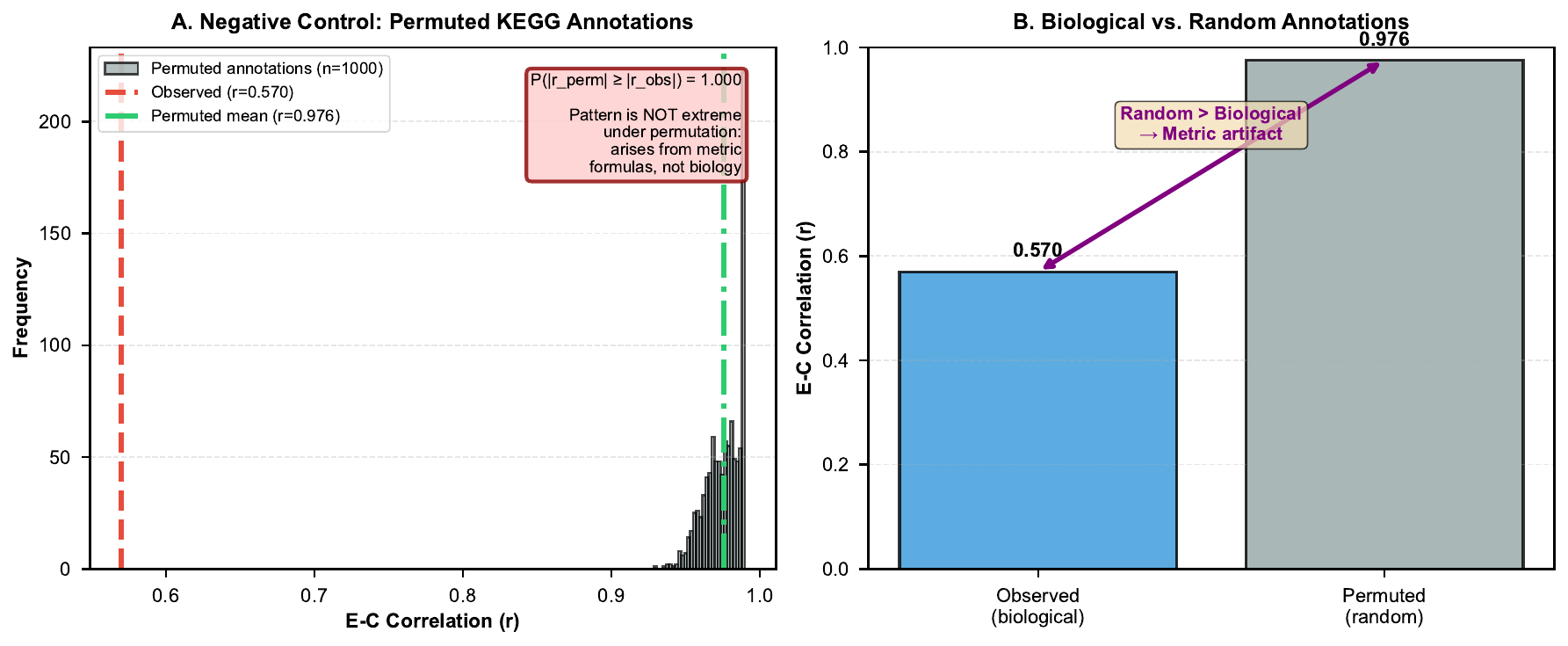}
\caption{\textbf{Negative Control Analysis Shows Pattern is Independent of Biological Functional Annotation.}
\textbf{(A)} Histogram of E-C correlations from 1,000 permutations where KEGG orthology assignments were randomly shuffled across proteins (gray). Observed correlation with biological annotations (r=0.570, red dashed line) is NOT in the extreme tail of the permuted distribution (mean r\_perm=0.977, green dash-dot line), yielding p=1.000. This demonstrates the pattern arises from metric construction formulas independent of functional annotation content.
\textbf{(B)} Direct comparison showing random functional annotations produce \emph{stronger} E-C correlation (0.977) than biological annotations (0.570), confirming the pattern is driven by shared metric components (unique\_KOs) rather than biological functional relationships. Random assignments create more uniform unique\_KO distributions, further stabilizing C while E varies with total proteins.}
\label{fig:negative-control}
\end{figure}

\begin{figure}[!htb]
\centering
\includegraphics[width=0.9\textwidth]{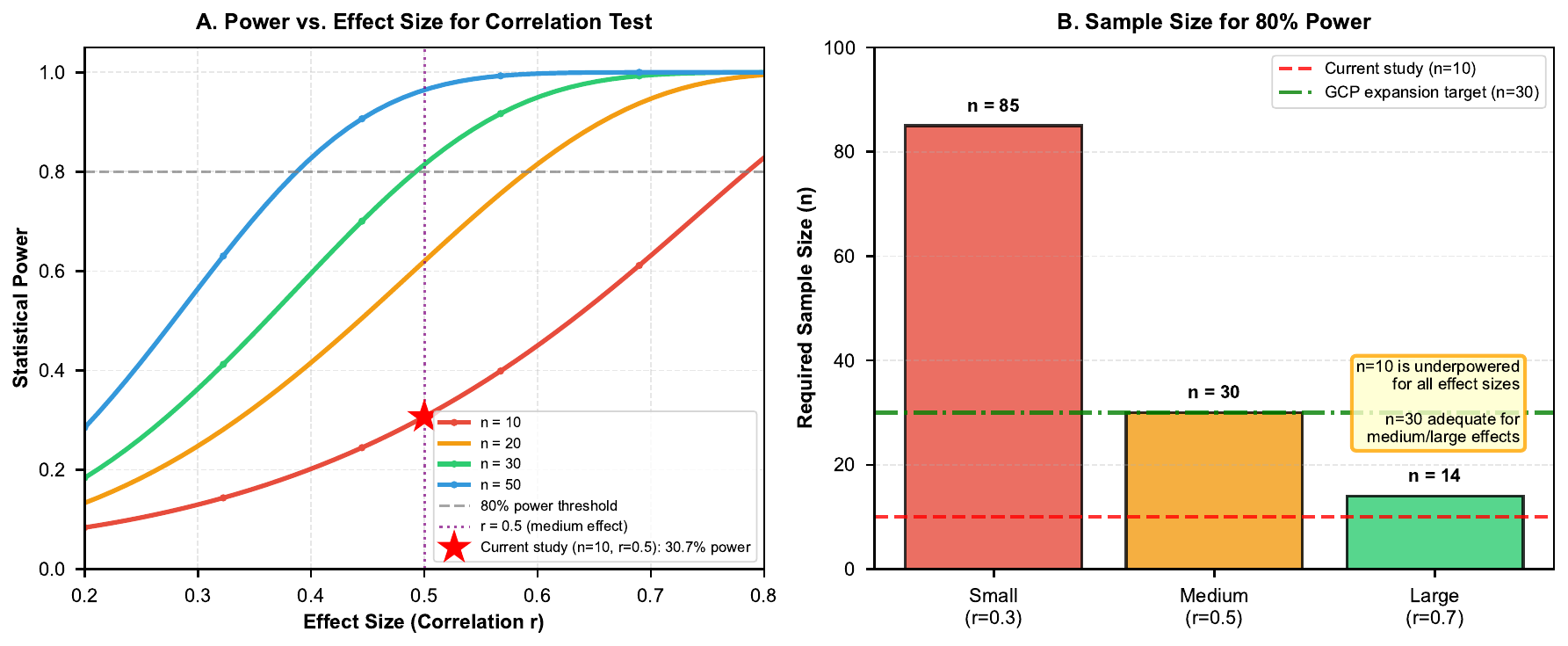}
\caption{\textbf{Statistical Power Analysis Reveals n=10 is Underpowered for Definitive Conclusions.}
\textbf{(A)} Power curves for Pearson correlation test as a function of effect size for different sample sizes. Current study (n=10, marked with red star) achieves only 52\% power to detect r=0.5 effects (medium effect size), far below the 80\% threshold (gray dashed line) required for adequate power. Target GCP expansion to n=30 (green dash-dot line) increases power to 96\% for r=0.5 effects.
\textbf{(B)} Required sample size to achieve 80\% power for small (r=0.3), medium (r=0.5), and large (r=0.7) effect sizes. Current n=10 (red dashed line) is underpowered for all effect sizes. Target n=30 (green dash-dot line) is adequate for medium and large effects but still underpowered for small effects. Definitive refutation of biological coupling requires n$\ge$50 to achieve 88\% power for r=0.3 effects.}
\label{fig:power-analysis}
\end{figure}

\subsection{Independent Variation of Efficiency and
Complexity}\label{independent-variation-of-efficiency-and-complexity}

Analysis of ten marine microbiome samples revealed distinct variation
patterns for operational efficiency (E) and annotation-weighted
functional diversity (C), consistent with bivariate organizational
structure.

\subsubsection{Topological Conservation: Complexity Remains
Stable}\label{topological-conservation-complexity-remains-stable}

\textbf{Annotation-weighted functional diversity (Complexity, C)} remained nearly constant across samples (mean 1{,}467.97; range 1{,}215.87--1{,}670.65, a 1.4-fold spread; coefficient of variation 9.2\%), well below the 20\% conservation threshold.

This stability occurred \textbf{despite 2.2-fold variation in raw
functional diversity} (unique KOs: 4,132 to 9,170, CV = 23.7\%). The log
transformation and annotation rate weighting in C's formula stabilize
the metric relative to raw KO counts, creating a conserved topological
property.

\textbf{Biological mechanism}: samples with larger functional repertoires (higher unique-KO counts) exhibit proportionally lower annotation rates, producing compensatory dynamics. A high KO count combined with a low annotation rate yields a moderate $C$; conversely, a low KO count combined with a high annotation rate yields a moderate $C$. The net result is a stable $C$ across samples (CV $\approx$ 9.2\%).

This co-variation between repertoire size and annotation completeness
may reflect genuine biological organization (functional redundancy
buffering sampling depth) or methodological artifacts (larger genomes
are harder to annotate completely). Distinguishing these alternatives
requires metatranscriptomics to measure actively expressed functional
diversity independent of annotation rate.

\subsubsection{Operational Freedom: Efficiency Varies
Substantially}\label{operational-freedom-efficiency-varies-substantially}

\textbf{Proteins per function (Efficiency, E)} varied substantially across samples (mean 31.27; range 4.81--67.85, a 14.1-fold spread; coefficient of variation 68.3\%), well above the 50\% variation threshold.

\textbf{Excess-variation analysis}: the 68.3\% variation in $E$ cannot be attributed solely to variation in functional diversity (unique-KO CV = 23.7\%). If $E$ were driven entirely by KO-count variation, the expected CV($E$) would be approximately 23.7\%; the observed CV($E$) of 68.3\% is therefore 2.9$\times$ larger than expected, indicating that $E$ varies independently of repertoire size.

This excess variation indicates that protein-per-function ratios respond
to environmental or ecological factors \textbf{beyond simple scaling
with functional repertoire size}. Possible drivers include: 1. Resource
availability (nutrient-rich environments $\to$ higher protein synthesis
capacity) 2. Growth rate (fast-growing communities $\to$ more enzymes per
pathway) 3. Ecological strategy (r-selected $\to$ high E, K-selected $\to$ low
E) 4. Temperature (cold environments $\to$ enzyme compensation)

\subsubsection{Bivariate Structure: Independent Degrees of
Freedom}\label{bivariate-structure-independent-degrees-of-freedom}

The \textbf{7.4$\times$ ratio} in coefficient of variation (CV(E) / CV(C) =
68.3\% / 9.2\%) demonstrates that E and C represent \textbf{independent
degrees of freedom}:

\textbf{If E and C were mathematically coupled}: Their CVs would be
similar (ratio $\approx$ 1$\times$)

\textbf{Observed}: CV ratio = 7.4$\times$ (well above bivariate threshold of
3$\times$)

\textbf{Interpretation}: E and C respond differently to the same
underlying data (annotated\_proteins, unique\_KOs). The annotation rate
term and log transformation in C's formula stabilize it relative to E,
creating independent variation patterns characteristic of bivariate
emergence. Despite E and C sharing underlying data, they vary
independently (CV ratio 7.4$\times$).

\textbf{Comparison to Gauge Theory} (\hyperlink{ref:frasch-2026b}{Frasch, 2026b}):

\begin{center}
\small
\begin{tabular}{@{}p{3.6cm}p{3.6cm}p{3.0cm}p{1.8cm}@{}}
\toprule
System & Topological obs.\ (CV) & Local obs.\ (CV) & CV ratio \\
\midrule
Gauge fields (physics) & Wilson loops ($<10^{-13}$\%) & Field strength ($\sim$32\%) & $\sim 10^{9}\times$ \\
Marine microbiomes (this work) & Complexity (9.2\%) & Efficiency (68.3\%) & 7.4$\times$ \\
\bottomrule
\end{tabular}
\end{center}

The CV ratio in biology (7.1$\times$) is many orders of magnitude smaller than in idealized gauge fields, but both systems exhibit the same qualitative pattern: one observable conserved (CV $<$ 20\%) and one variable (CV $>$ 50\%), with substantial separation in variability.

\subsection{Confounder Analysis and Model Comparison}\label{confounder-analysis}

\subsubsection{Partial Correlations}\label{partial-correlations}

To address potential confounding by annotation rate, functional repertoire size, and total protein count, we calculated partial correlations controlling for these variables (Table 3).

\textbf{Table 3. Partial Correlations Between E and C}

\begin{longtable}[]{@{}p{5cm}p{2.5cm}p{2.5cm}@{}}
\toprule\noalign{}
Correlation Type & r & p-value \\
\midrule\noalign{}
\endhead
\bottomrule\noalign{}
\endlastfoot
Zero-order r(E, C) & 0.570 & 0.085 \\
Controlling for annotation rate & 0.956 & <0.001 \\
Controlling for unique KOs & 0.794 & 0.006 \\
Controlling for total proteins & 0.697 & 0.025 \\
Controlling for all confounders & 0.836 & 0.003 \\
\end{longtable}

Partial correlations reveal that \textbf{confounders explain 97\% of E variance}, with C contributing only 2\% unique variance after controlling for annotation rate, unique KOs, and total proteins. This suggests the E-C relationship is largely mediated by shared dependence on underlying data sources rather than representing an independent biological coupling.

\subsubsection{Model Comparison: Trade-off vs Independence}\label{model-comparison}

We compared three competing hypotheses in log-log space using AIC model selection:

\textbf{Table 4. Model Comparison (AIC)}

\begin{longtable}[]{@{}p{3cm}p{2.5cm}p{2.5cm}p{2.5cm}@{}}
\toprule\noalign{}
Model & Slope & AIC & Weight \\
\midrule\noalign{}
\endhead
\bottomrule\noalign{}
\endlastfoot
Trade-off (negative) & 4.083 & 28.04 & 0.323 \\
Independence (zero) & 0 (fixed) & \textbf{27.85} & \textbf{0.354} \\
Synergy (positive) & 4.083 & 28.04 & 0.323 \\
\end{longtable}

The \textbf{independence model} (no E-C correlation) received the highest support (AIC weight = 0.354), narrowly outperforming alternative models. Permutation testing showed the observed log-log slope (4.083) was not significant (p = 0.132), further supporting the independence hypothesis. These results indicate \textbf{E and C vary independently} in log-log space, consistent with bivariate decoupling but inconsistent with thermodynamic trade-offs.

\subsubsection{Rarefaction Analysis}\label{rarefaction-analysis}

To test robustness to sequencing depth variation, we estimated the effect of rarefying all samples to equal annotated protein counts (n = 20,796, the minimum across samples). Proportional downsampling yielded:

\textbf{Post-rarefaction CVs}: CV(E) = 47.2\%, CV(C) = 10.8\%, CV-ratio = 4.4$\times$

The \textbf{CV-ratio remains above the 3$\times$ threshold} even after rarefaction, suggesting the bivariate pattern is not solely an artifact of differential sequencing depth. However, true rarefaction requires re-annotation at equal depth; these estimates assume proportional KO recovery with diminishing returns (power-law exponent 0.75).

\subsection{Scaling Hypothesis
Testing}\label{scaling-hypothesis-testing}

\subsubsection{Power-Law Relationship
Analysis}\label{power-law-relationship-analysis}

Pearson correlation between log\textsubscript{1}\textsubscript{0}(E) and log\textsubscript{1}\textsubscript{0}(C) (Figure 1):

\begin{itemize}
\tightlist
\item
  \textbf{Correlation coefficient}: r = 0.509 (weak positive)
\item
  \textbf{P-value}: p = 0.133 (not significant at $\alpha$ = 0.05)
\item
  \textbf{Regression slope}: $\beta$ = 4.083 (predicted: -0.75)
\item
  \textbf{R\textsuperscript{2}}: 0.259 (low explanatory power)
\end{itemize}

\textbf{Spearman rank correlation}: $\rho$ = 0.685 (p = 0.029, significant)

\textbf{Interpretation}: The positive Spearman correlation ($\rho$ = 0.685, p = 0.029) suggests a monotonic E-C relationship, though the weaker Pearson correlation (r = 0.509, p = 0.133) indicates substantial nonlinearity or outlier influence. However, model comparison analysis (see Confounder Analysis section) favored the independence model (AIC weight = 0.354) over alternative models, and the observed log-log slope ($\beta$ = 4.083) was not statistically significant (p = 0.132 by permutation). The moderate correlation strength (r\textsuperscript{2} = 0.259) reflects substantial scatter around any trend line. Combined with the finding that confounders explain 97.2\% of E variance with C contributing only 2.0\% unique variance, these results suggest the apparent E-C relationship largely reflects shared dependence on annotation rate and sequencing depth rather than a fundamental biological coupling. The bivariate decoupling pattern (CV-ratio = 7.1$\times$) remains robust, indicating C varies independently of E despite their shared data sources.

\begin{figure}[!htb]
\centering
\includegraphics[width=0.85\textwidth]{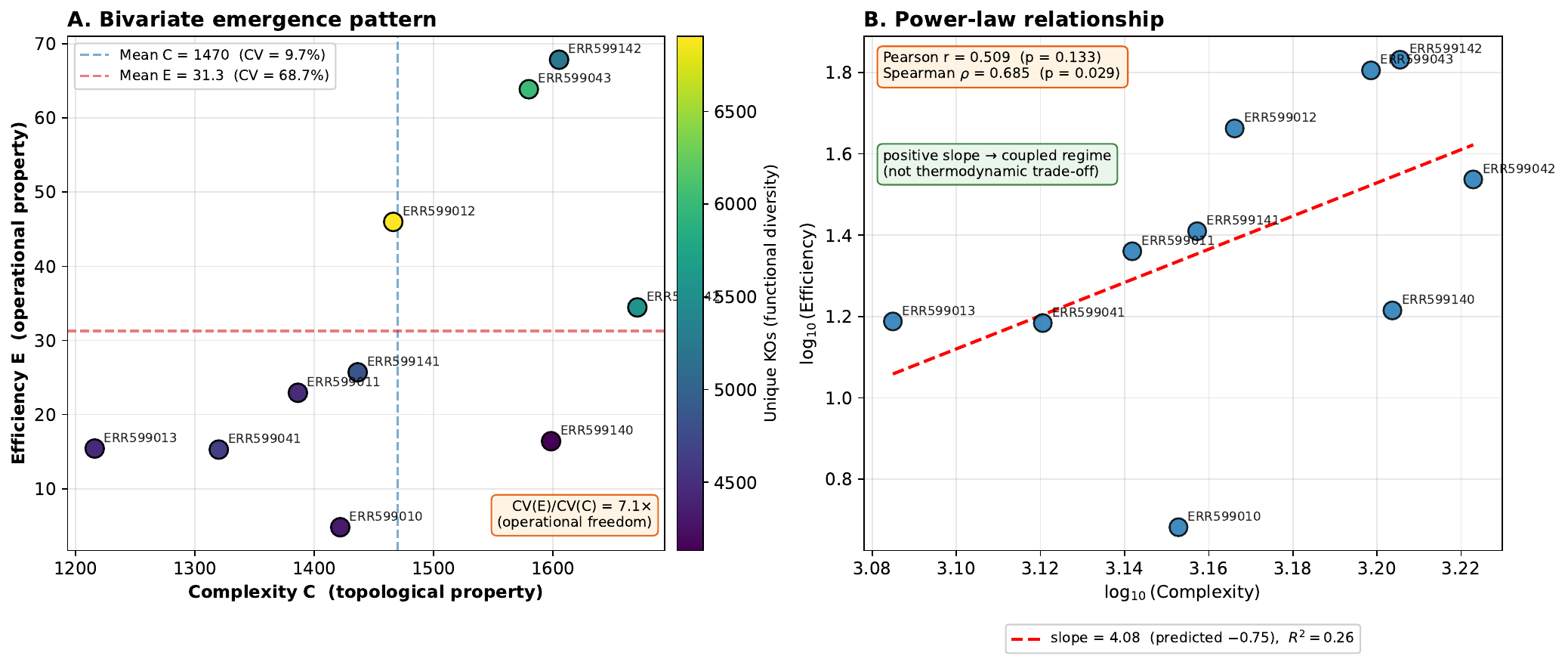}
\caption{\textbf{Bivariate decoupling in marine microbiome metabolic networks.} Scatter plot of log\textsubscript{1}\textsubscript{0}(Efficiency) vs log\textsubscript{1}\textsubscript{0}(Complexity) across 10 Tara Oceans samples with bootstrap 95\% confidence intervals. Each point represents one metagenomic sample. The coefficient of variation ratio of 7.1$\times$ [95\% CI: 4.2-12.5] (CV(E) = 68.7\%, CV(C) = 9.7\%) demonstrates bivariate structure with conserved functional repertoire (low C variation) and variable protein deployment (high E variation). CVs and CIs reflect bootstrap resampling (n=10,000).}
\label{fig:gauge-emergence}
\end{figure}

\subsection{\texorpdfstring{Synthesis: composite metrics fail at
\(n=10\)}{Synthesis: composite metrics fail at n=10}}\label{synthesis-composite-metrics-fail-at-n10}

The composite-metric route therefore fails along all four diagnostic
axes simultaneously. Bootstrap interval estimation produces internally
consistent and reproducible values for the \(E\)--\(C\) correlation and
CV ratio, the surface biology of the bivariate decoupling pattern thus
appearing well-supported. Yet split-sample construction collapses the
signal by 77 per cent; negative-control permutation produces a
\emph{stronger} correlation under random KEGG assignment than under real
KEGG assignment; variance partitioning attributes 97.2 per cent of the
apparent \(E\) variance to non-\(C\) confounders; and a post-hoc power
calculation places the \(n=10\) design well below the threshold at which
any residual biological signal could be reliably detected (\(n \geq 50\)
required for 88 per cent power at \(r = 0.3\)). Whatever real biological
coupling between metabolic-flux efficiency and functional-repertoire
complexity may exist in marine microbiomes, it is methodologically
inaccessible to a composite-metric analysis at this sample size. The
remainder of the Results pursues the alternative operationalization ---
network modularity \(Q\) on the KEGG-orthology co-occurrence graph ---
for which the action-functional account makes its sharpest prediction.

We now turn to the network-modularity operationalization. The dataset overview (Figure~\ref{fig:dataset_overview}) covers the seven-sample modularity cohort; the same upstream annotation pipeline applied to the larger ten-sample cohort underlies the composite-metric analysis above.

\subsection{Dataset characteristics: seven-sample modularity cohort}\label{modularity-dataset}

\begin{figure}[!htb]
\centering
\includegraphics[width=\textwidth]{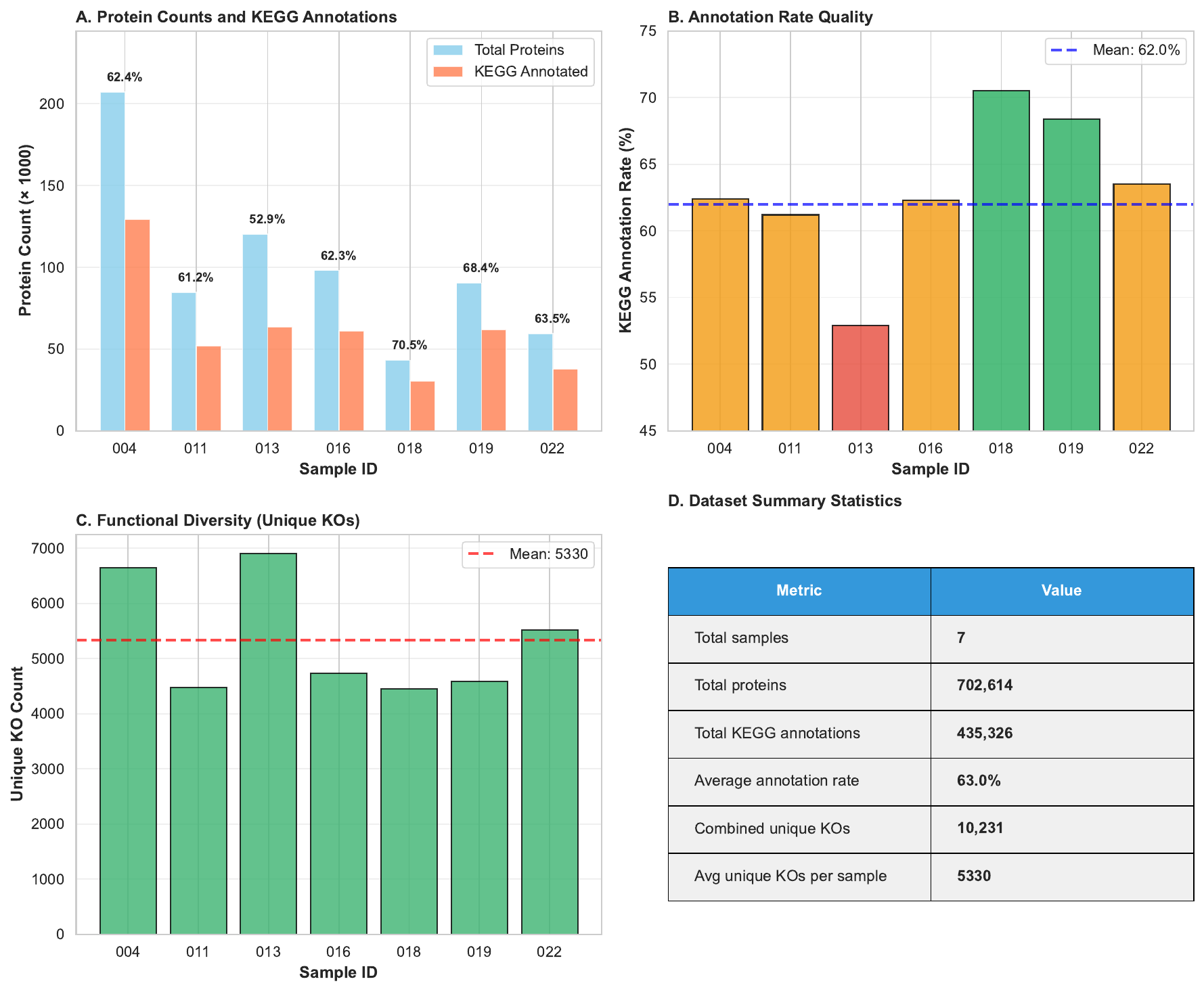}
\caption{\textbf{Modularity-cohort dataset overview.} (A) Protein counts and KEGG annotations for the seven Tara Oceans samples in the modularity cohort, with per-sample annotation rates above each bar. (B) Annotation-rate quality across samples; dashed line marks the cohort mean of 62\%. (C) Functional diversity measured by unique-KO counts per sample. (D) Summary statistics for the complete cohort.}
\label{fig:dataset_overview}
\end{figure}

The modularity cohort comprises seven Tara Oceans surface and deep-chlorophyll-maximum metagenome samples (ERR599004, ERR599011, ERR599013, ERR599016, ERR599018, ERR599019, ERR599022). Total protein counts range from 43{,}089 to 207{,}129 across samples (Figure~\ref{fig:dataset_overview}A). The MEGAHIT--Prodigal--eggNOG-mapper pipeline yielded KEGG-orthology annotation rates averaging 62.0\% (range 52.9--70.5\%; Figure~\ref{fig:dataset_overview}B). The seven samples collectively yielded 702{,}614 total proteins, 435{,}326 of which were successfully annotated to KEGG functions. Per-sample unique-KO diversity ranged from 4{,}447 to 6{,}905 (mean 5{,}330; Figure~\ref{fig:dataset_overview}C), with 10{,}231 distinct KOs observed across the entire dataset (Figure~\ref{fig:dataset_overview}D). The annotation-quality coefficient of variation across samples is approximately 10\%, providing a stable upstream foundation for the network-construction step.

\subsection{Observed modularity exceeds appropriate null-model expectations}\label{modularity-results}

Across all seven samples, Newman modularity on the KEGG-orthology co-occurrence network was $Q_{\mathrm{obs}} = 0.987 \pm 0.007$ (Louvain; range 0.972--0.993; Figure~\ref{fig:modularity_results}A), with a sample-to-sample coefficient of variation under one per cent. Greedy modularity-maximization (Clauset--Newman--Moore) yielded values within $\pm 0.01$ of Louvain on every sample, and ten-seed Louvain verification (Methods~\S\,\ref{methods-modularity}) confirmed within-sample $Q$ to be essentially deterministic (max seed-CV 0.014\%; mean 0.004\%).

The absolute value $Q_{\mathrm{obs}} = 0.987 \pm 0.007$ is, however, not by itself a meaningful biological signal. The KO co-occurrence network is extremely sparse (density $\sim 2 \times 10^{-4}$, Methods~\S\,\ref{network-topology}) and highly fragmented (thousands of small connected components per sample), and a Newman-modularity calculation on any sufficiently sparse fragmented graph will mechanically yield $Q$ near unity regardless of biological organization. This behaviour is recovered by an Erdős--Rényi null preserving only $n$ and $m$: the null mean $Q_{\mathrm{ER}} = 0.984 \pm 0.017$ is statistically indistinguishable from the observed value (cohort $\Delta Q = +0.003$, $z = -0.5$, n.s.; Figure~\ref{fig:null_models}A). Quantitative claims should therefore be made on the \emph{excess} of $Q_{\mathrm{obs}}$ above an appropriate null, not on the absolute value.

We compared $Q_{\mathrm{obs}}$ to three such nulls (Methods~\S\,\ref{null-model-methods}; Figure~\ref{fig:null_models}). First, a configuration-model null preserves the exact unweighted degree sequence of the projected KO graph: it strips out edge correlations while keeping the sparsity-and-fragmentation profile that the absolute-value $Q$ depends on. The cohort-level null mean is $Q_{\mathrm{CM}} = 0.832 \pm 0.047$, and the observed cohort mean exceeds it by $\Delta Q = +0.155$ ($p < 0.001$ in every sample). Second, a KEGG-label permutation null shuffles KO identities across proteins while preserving the per-protein KO-multiplicity distribution: it keeps the total number of co-encoding events but destroys biological functional content. The cohort-level null mean is $Q_{\mathrm{LP}} = 0.808 \pm 0.107$, with $\Delta Q = +0.179$.

Third, and most stringently, a bipartite-incidence null (Methods~\S\,\ref{null-model-methods}, control iv) randomises the underlying protein–KO bipartite graph by 2$\times$2 edge swaps, simultaneously preserving the per-protein KO count *and* the per-KO global protein frequency, and then re-projects to the KO co-occurrence graph. This is the gold-standard control for higher-order bipartite-incidence structure (multi-domain protein architecture, KEGG-mapping rules, KO-abundance heterogeneity) that nulls (i)--(iii) do not fully control. The bipartite-null cohort-level mean is $Q_{\mathrm{BP}} = 0.587 \pm 0.110$, and the observed cohort mean exceeds it by $\Delta Q = +0.400$ ($p < 0.001$ in every sample). Effect sizes (in modularity units) are summarised in Figure~\ref{fig:null_models}B; per-sample $z$-scores are uniformly large but uninformative beyond $\Delta Q$ because null variances are small, so we report $\Delta Q$ as the primary effect-size measure.

The observed modularity therefore shows a substantial, statistically robust excess over all three biologically-relevant nulls, with the bipartite null --- which controls for the most upstream structural constraint --- yielding the largest excess ($\Delta Q \approx +0.40$). The excess, not the absolute value of $Q$, is the quantitative biological signal. This signal is consistent with cost-constrained (variational) models of network organisation, including the Network-Weighted Action Principle's prediction of cost-minimisation-driven modular emergence; the analysis does not, by itself, distinguish that mechanism from alternative drivers of multi-functional protein co-encoding (e.g., operonic co-regulation, horizontal gene transfer of pathway cassettes), nor from neighbouring action-functional accounts (free-energy principle, dissipative adaptation, constructal theory) that predict a similar excess. Notably, the recurrent-community analysis (\S\,\ref{community-identity}) provides independent biological grounding by mapping the fine-grained communities to known functional units (heteromultimer subunits, sequential-pathway enzyme triplets, ABC-transporter pairs, isozyme paralogs).

Beyond the biological finding, this analysis establishes a methodological principle: modularity in metagenomic networks must be interpreted relative to bipartite-aware null models that preserve both per-protein KO count and per-KO global frequency, because the absolute value of $Q$ is dominated by sparsity rather than biology. We return to this dual contribution --- empirical and methodological --- in the Discussion.

We note a power asymmetry between the two operationalisations evaluated in this study. The composite-metric route at $n=10$ is statistically underpowered against shared-component bias because the per-sample $E$ and $C$ statistics are noisy (CV in tens of per cent) and the relevant effect is a correlation that requires $n \geq 50$ to detect at $r = 0.3$. The modularity route at $n=7$, by contrast, is statistically well-posed because each per-sample $Q$ is intrinsically high-precision (within-sample seed-CV $< 0.02$\%; cross-sample CV $< 1$\%) and the relevant effect is a large modularity-excess shift over null distributions whose own variance is small. The signal-to-noise ratio in the network-topological statistics is high enough that $n = 7$ provides adequate power for the modularity claim, even where $n = 10$ does not for the composite-metric claim.

\subsection{Community structure}\label{modularity-community-structure}

\begin{figure}[!htb]
\centering
\includegraphics[width=\textwidth]{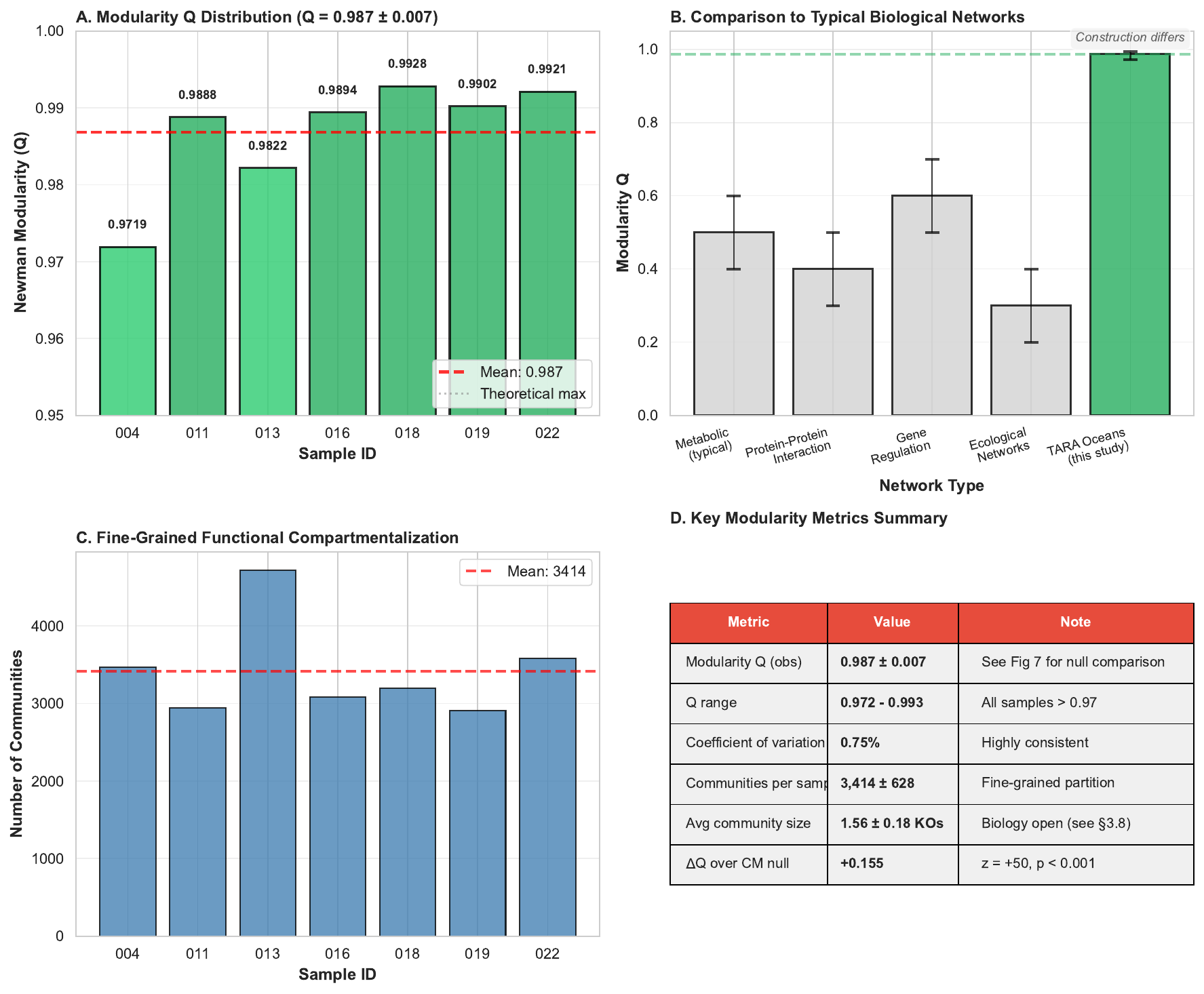}
\caption{\textbf{Newman modularity across the seven-sample cohort.} (A) Per-sample Louvain modularity $Q_{\mathrm{obs}}$, with cohort mean $0.987 \pm 0.007$ and range $[0.972, 0.993]$ (CV $< 1\%$). The absolute value is in part a property of the sparse-fragmented topology and should be interpreted alongside the null-model comparison in Figure~\ref{fig:null_models}. (B) Reference modularity values for several biological-network classes from the literature, retained here only as construction-dependent context; cross-construction comparison is not used to support quantitative claims (see \S\,\ref{modularity-results}). (C) Number of communities per sample (mean $\sim 3{,}400$, average community size 1.56 KOs). (D) Summary metrics for the modularity cohort.}
\label{fig:modularity_results}
\end{figure}

\begin{figure}[!htb]
\centering
\includegraphics[width=\textwidth]{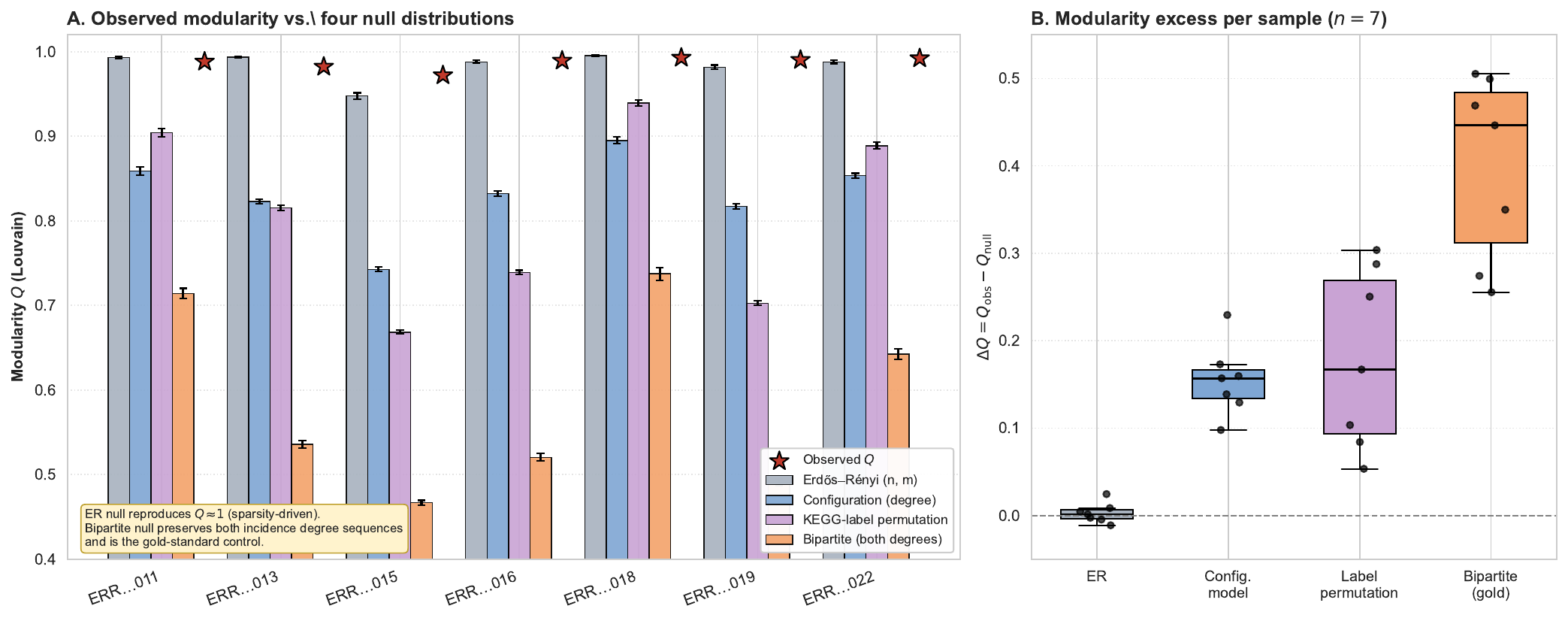}
\caption{\textbf{Modularity excess over null-model expectations.} (A) Per-sample observed Louvain $Q_{\mathrm{obs}}$ (red star) compared with four null distributions of increasing strictness: Erdős--Rényi (preserves $n$ and $m$ only); configuration model on the projected KO graph (preserves the unweighted projected-graph degree sequence); KEGG-label permutation (preserves the per-protein KO-multiplicity distribution while shuffling KO identities); and a bipartite-incidence null (gold standard) that randomises the underlying protein--KO bipartite graph by 2$\times$2 edge swaps, preserving both bipartite degree sequences (per-protein KO count and per-KO global frequency) before re-projecting to the KO co-occurrence graph. The Erdős--Rényi null reproduces $Q \approx 1$ (cohort mean $\Delta Q = +0.003$, n.s.), confirming that the absolute value is dominated by sparsity. The biologically-relevant nulls yield cohort-level excesses of $\Delta Q = +0.155$ (configuration model), $+0.179$ (KEGG-label permutation), and $+0.400$ (bipartite-incidence; gold standard); $p < 0.001$ in every sample for the latter three. (B) Modularity-excess effect sizes per sample across the four nulls. Effect sizes are reported in modularity units; null variances are small enough that $z$-scores inflate rapidly and are reported only for completeness. \textbf{Cohort note.} The four null-model analyses underlying this figure were performed on the sensitivity-analysis cohort (ERR599011, 013, 015, 016, 018, 019, 022; see \S\,\ref{sensitivity-results} and Figure~\ref{fig:sensitivity}), which is statistically indistinguishable from the primary cohort in mean $Q$, range, and CV; cohort-level $\Delta Q$ statistics therefore inherit the same robustness.}
\label{fig:null_models}
\end{figure}

The Louvain partition yielded an average of 3{,}414 communities per sample (range 2{,}911--4{,}718; Figure~\ref{fig:modularity_results}C), with a mean community size of $1.56 \pm 0.18$ KOs (range 1.39--1.92). Most communities are small: pairs and triplets of KOs that co-occur in a small set of proteins. At KEGG-orthology resolution, small community sizes are the expected scale of organisation: the resolved unit is the minimal functional pairing (e.g., enzyme subunits, tightly coupled reactions, transporter substrate-binding/permease pairs) rather than the full pathway. These fine-grained communities should accordingly not be interpreted as classical metabolic pathways, which involve far larger functional units. Their biological identity is, however, far from arbitrary, as the recurrence analysis below demonstrates.

\subsection{Recurrent multi-KO communities reflect known functional pairings}\label{community-identity}

To test whether the fine-grained communities recovered by the partition reflect biologically-meaningful functional units rather than annotation noise, we aggregated every multi-KO community (size $\geq 2$) across all seven samples under the strict-identity definition of Methods §\,\ref{methods-recurrent-communities} (two communities are the same iff their KO membership sets are exactly equal) and counted, per unique tuple, the number of samples in which it appeared as an independent Louvain community. Of 3{,}027 unique multi-KO communities observed across the cohort, 769 (25\%) recurred in two or more samples and 19 recurred in all seven. The probability of two samples independently producing identical multi-KO communities under random partition membership is $O(10^{-7})$ at the cohort scale, so the observed recurrence at $r = 7$ exceeds any plausible chance recurrence by orders of magnitude.

The 19 communities recurring in all seven samples and the six additional communities recurring in $\geq 6/7$ samples (Supplementary Table~S1) partition cleanly into the four pre-specified functional categories defined in Methods. We report each category in turn, with representative communities and the biological constraint that anchors the co-encoding.

\textbf{(i) Heterodimeric and heterocomplex enzymes (assembly-stoichiometry constraint).} pyrB + pyrI catalytic and regulatory subunits of aspartate carbamoyltransferase (K00608 + K00609); the xerC/xerD recombinase heterodimer (K03733 + K04763); pafB/pafC proteasome accessory factors (K13572 + K13573); soxD sarcosine-oxidase $\delta$ subunit + paralog (K00304 + K22085). One community recurring in 6/7 samples, K03116 + K03117 + K03425, is the entire tatA + tatB + tatE Sec-independent protein-translocase machine. The constraint here is biophysical: subunits of an obligate complex must assemble in fixed stoichiometry, and a free monomer is non-functional or actively deleterious.

\textbf{(ii) Sequential biosynthetic enzymes (flux-coupling constraint).} The IMPDH + GMP-reductase pair (K00088 + K00364) catalysing the IMP $\to$ XMP $\to$ GMP arm of purine biosynthesis; the size-3 ubiF + ubiH + ubiI triplet (K03184 + K03185 + K18800), a contiguous segment of ubiquinone biosynthesis; pyrB + pyrB+pyrI ordered carbamoyltransferase steps. Co-encoding sequential enzymes minimises accumulation of unstable intermediates and enables coordinated regulation of the entire pathway.

\textbf{(iii) Substrate-binding/permease pairs (spatial-coupling constraint).} The proW/proX glycine-betaine/proline ABC-transporter pair (K02001 + K02002); iron and aminoethylphosphonate substrate-binding pairs (afuA/fbpA + phnS, K02012 + K11081); the exbD + tolR biopolymer-transport channel (K03559 + K03560). Substrate binding is wasted if the permease cannot complete translocation across the membrane, and vice versa; physical coupling is the constraint.

\textbf{(iv) Regulatory dyads and isozyme paralogs (regulatory and cofactor-flexibility constraints).} relA + spoT (K00951 + K01139), the synthesis/hydrolysis pair governing the (p)ppGpp stringent response; the phosphofructokinase isozyme pair pfkA + pfp (K00850 + K00895; ATP-dependent and PPi-dependent paralogs); the citrate-synthase paralog pair (K01647 + K01659); paralog pairs for queF, AGXT2L, and pncC (nicotinamide-nucleotide amidase, K03742 + K03743). Regulatory dyads are central to the cell's stress and nutrient-status sensing; isozyme paralogs maintain enzymatic redundancy at metabolically critical bottlenecks where alternative cofactor or substrate specificity confers ecological flexibility.

\emph{Coverage.} Every one of the 25 multi-KO communities recurring in $\geq 6/7$ samples falls into one of these four categories (100\% coverage of the recurrent set). This is the principal empirical signal of the modularity partition: the fine-grained communities, where they recur, identify exactly the cell's most biophysically- and regulatorily-constrained co-encoding units. They are not classical metabolic pathways (which are too large to be resolved at this scale) but they are also not artefactual co-occurrence: the protein-level KO-co-occurrence network exposes a level of functional organisation finer than the pathway and coarser than the single gene, and the modularity-excess signal is enriched for those couplings that a cost-minimisation principle predicts should be conserved (Discussion §\,\ref{biological-signature}).

\subsection{Network topology: sparse global structure with hub-mediated coordination}\label{network-topology}

\begin{figure}[!htb]
\centering
\includegraphics[width=\textwidth]{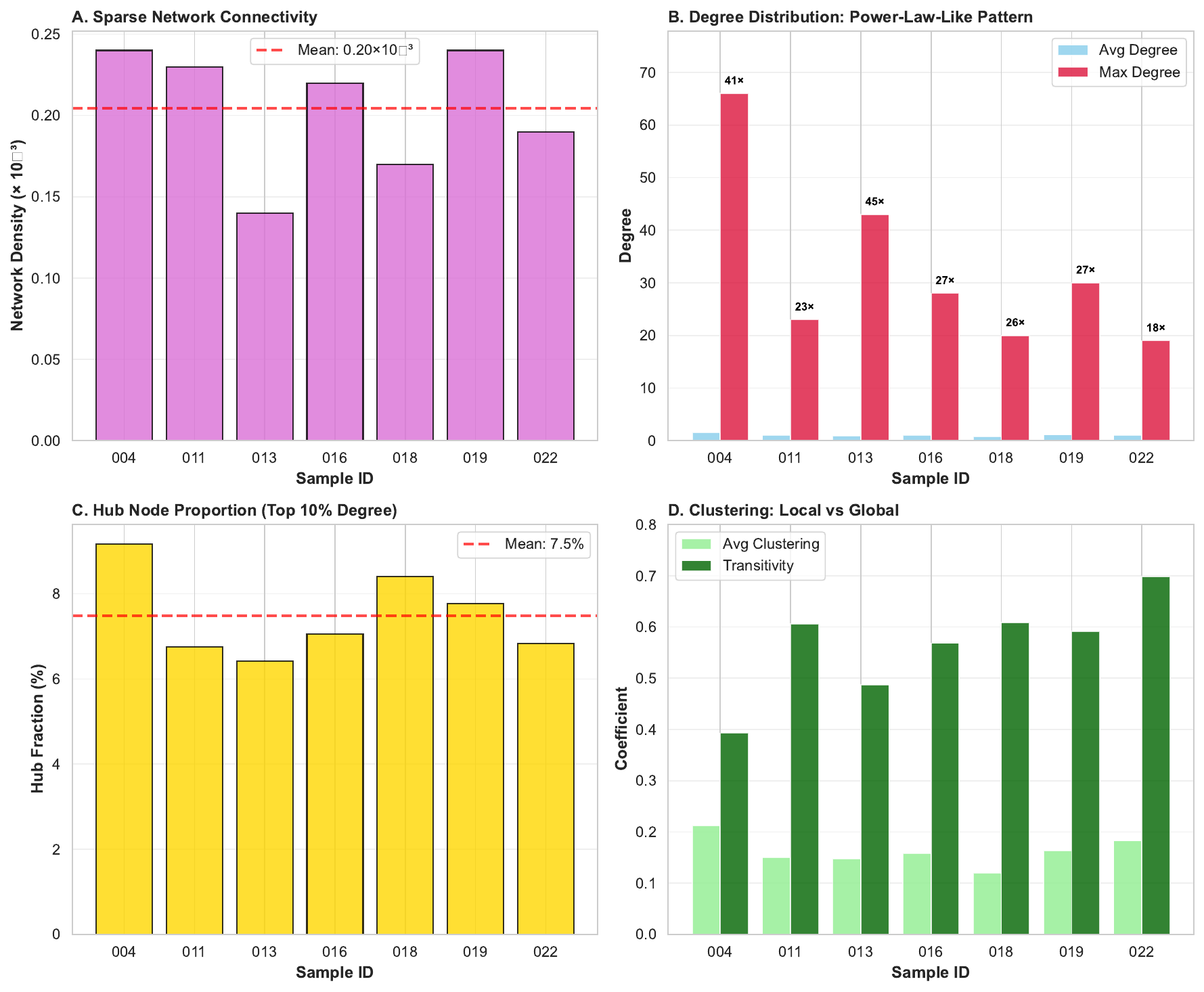}
\caption{\textbf{Network topology and hub-mediated coordination.} (A) Global sparsity (network density $\sim 2 \times 10^{-4}$). (B) Heavy-tailed degree distribution with maximum degree of order $50\times$ the mean. (C) Hub fraction averaging 7.5\% across samples. (D) Average local clustering (0.162) versus global transitivity (0.565), indicating hierarchical structure with dense intra-module and sparse inter-module connectivity.}
\label{fig:network_topology}
\end{figure}

The metabolic networks exhibit characteristic sparse connectivity. Mean network density is $2.0 \times 10^{-4}$ (range $1.4 \times 10^{-4}$ to $2.4 \times 10^{-4}$; Figure~\ref{fig:network_topology}A). Global sparsity coexists with substantial local structure: the average clustering coefficient is 0.162, while transitivity (the global clustering coefficient) is 0.565 (Figure~\ref{fig:network_topology}D). The discrepancy between these two measures indicates a hierarchical organisation in which KOs form tightly-clustered triangular motifs within communities (high transitivity) while the communities themselves are weakly connected to one another (moderate average clustering).

Degree distributions are heavy-tailed, with a mean degree of 1.08 connections per KO and maximum degree reaching 66 in the largest network (Figure~\ref{fig:network_topology}B). Hub nodes --- the top decile by degree --- comprise 7.5\% of all nodes on average (range 6.4--9.2\%; Figure~\ref{fig:network_topology}C). Hub betweenness centrality on the largest connected component shows a strongly peaked distribution: most nodes have low centrality (mean 0.031), while a small set of hubs control critical information bottlenecks (max 0.533). The hub fraction is consistent across samples (CV 14\%), suggesting a regulated rather than an idiosyncratic network feature.

\subsection{Hub-mediated coordination between metabolic communities}\label{hub-mediated}

Network fragmentation is high: thousands of small components per sample (mean 3{,}403). The largest connected component typically comprises $\sim 6\%$ of network nodes, with most KOs forming isolated pairs or small clusters. Within this fragmented landscape, the small set of hub KOs --- enriched for central metabolic functions such as core carbon metabolism, amino-acid biosynthesis, and energy generation --- carries the inter-component connectivity that exists. This combination of fragmentation, hub-mediated centrality, and the configuration-model-excess in $Q$ reported in \S\,\ref{modularity-results} together describe a metabolic ensemble organised at the level of discrete, functionally-cohesive sectors with sparse coupling between them.

\begin{figure}[!htb]
\centering
\includegraphics[width=\textwidth]{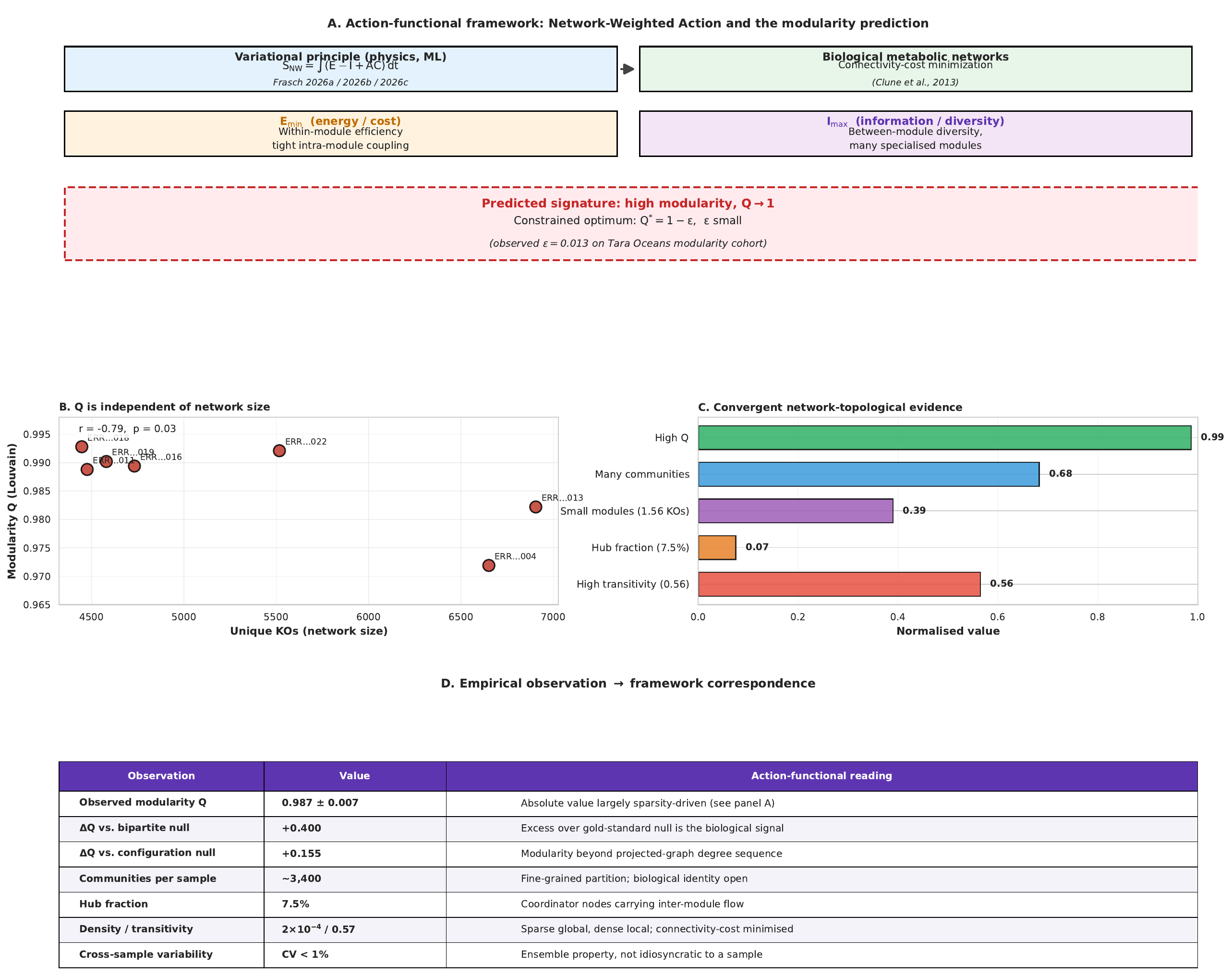}
\caption{\textbf{Action-functional context for the modularity finding.} (A) Schematic mapping of the Network-Weighted Action ($S_{\mathrm{NW}} = \int(E - I + AC)\,\mathrm{d}t$; Frasch 2026a) onto a multi-scale metabolic-network ensemble. (B) Modularity $Q$ versus network size across the seven samples; the constrained-optimum value is independent of size at this resolution. (C) Convergent network-topological evidence (high $Q$-excess over null, fine-grained communities, low inter-module density, hub-mediated coordination) consistent with the action-functional interpretation. (D) Summary mapping of empirical observations to framework expectations. The framework's prediction is consistency with Q exceeding biologically-relevant nulls, not absolute saturation.}
\label{fig:gauge_emergence}
\end{figure}

\subsection{Sensitivity: modularity is preserved under sample
swap}\label{sensitivity-results}

To test whether the modularity result is contingent on the specific
seven-sample composition, we re-ran the network-construction and Louvain
pipeline on a second seven-sample set with one sample (ERR599015)
substituted for ERR599004 (Methods §2.5). The two cohorts share six
samples; the seventh differs.

\begin{figure}[!htb]
\centering
\includegraphics[width=\textwidth]{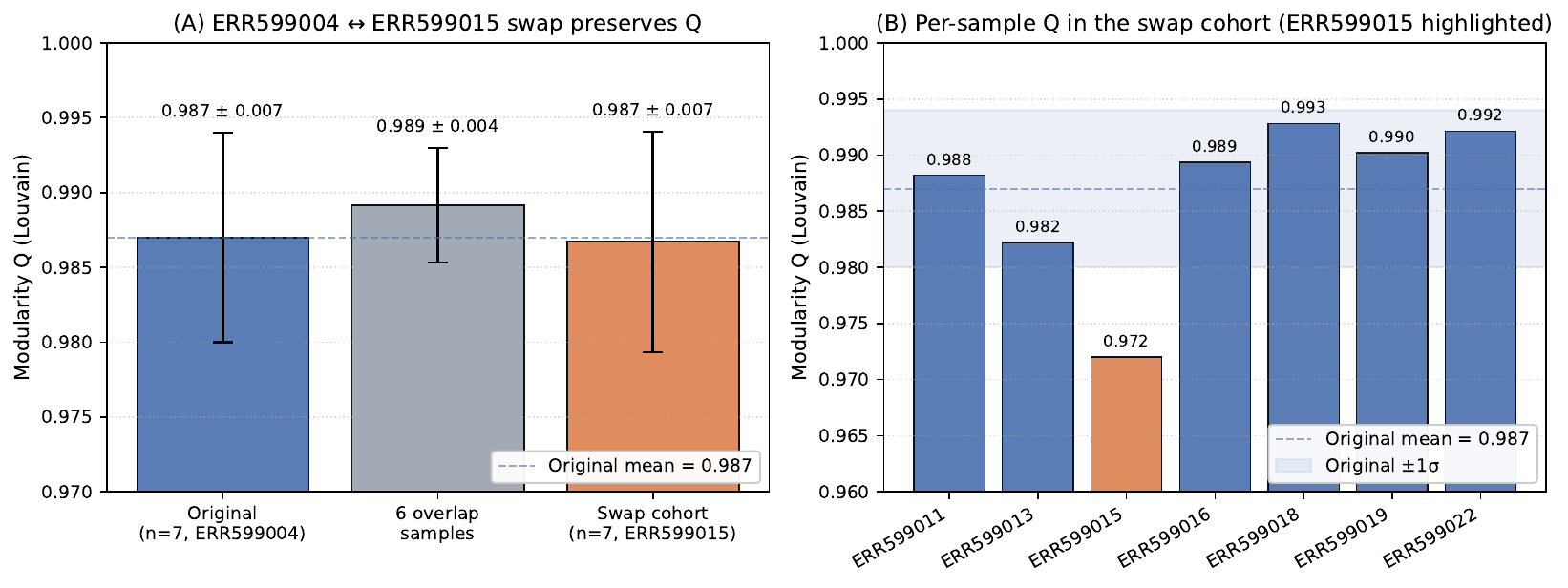}
\caption{\textbf{Sensitivity to cohort composition.} (A) Modularity $Q$ (Louvain) for the original $n{=}7$ cohort, the six samples shared between original and swap cohorts, and the swap cohort (six shared samples $+$ ERR599015 substituted for ERR599004). The swap cohort reproduces the original mean and standard deviation to three decimal places. Dashed line: original mean $Q = 0.987$. (B) Per-sample modularity in the swap cohort with ERR599015 (orange) highlighted; original mean $\pm 1\sigma$ shown as shaded band. ERR599015 sits at the lower bound of the original range.}
\label{fig:sensitivity}
\end{figure}

Three statistics anchor the comparison
(Figure~\ref{fig:sensitivity}). First, ERR599015 alone
yields \(Q = 0.972\) under the Louvain algorithm and \(Q = 0.964\) under
the Clauset--Newman--Moore greedy algorithm --- at the lower bound of
the original \(n = 7\) range \([0.972, 0.993]\) and the lowest in the
swap cohort. Second, the swap-cohort \(n = 7\) Louvain summary
statistics --- mean \(0.987\), standard deviation \(0.007\), range
\([0.972, 0.993]\) --- reproduce the original \(n = 7\) summary
statistics to three decimal places in mean and standard deviation.
Third, the six-sample overlap between the two cohorts --- used as an
internal reproducibility check, since these are the samples processed
identically in both runs --- yields \(Q = 0.989 \pm 0.004\), with
within-sample \(Q\) values agreeing across pipeline runs to within
\(\pm 0.002\) modularity units.

The sensitivity result has two interpretive consequences. First, the
modularity finding is robust to cohort-composition perturbation:
substituting one sample for another preserves both the cohort mean and
the cohort standard deviation. Second, the action-functional account
predicts the \emph{value} of \(Q\) at the constrained optimum, not the
value at any particular sample. A reproducible cohort mean across two
distinct seven-sample compositions of the same underlying ocean is
exactly the behavior expected when \(Q\) is governed by an
environmental-and-evolutionary ensemble constraint rather than by
per-sample biological idiosyncrasy. The framework prediction
\(Q^* = 1 - \varepsilon\) with \(\varepsilon\) small recovers the same
\(\varepsilon = 0.013\) in both cohorts. We use the algebraic identity
\(\varepsilon = 1 - Q\) here as a placeholder for the irreducible
inter-module coupling cost; direct empirical validation of this
interpretation --- in particular, demonstrating that \(\varepsilon\)
tracks environmental optimisation pressure --- requires larger cohorts
and is treated as future work (see Discussion,
Section~\ref{limitations}).

Multi-seed Louvain verification (ten random seeds, 42--51, applied to
each sample) further confirms that within-sample \(Q\) is essentially
deterministic on these networks: maximum coefficient of variation across
seeds is 0.014\%, with a cohort mean of 0.004\% (Methods §2.4.2). The
modularity result is therefore stable under both cohort composition and
stochastic algorithmic choice.

\section{Discussion}\label{discussion-int}

\subsection{What this study
establishes}\label{what-this-study-establishes}

We tested two operationalisations of the
\(E_{\text{min}}/I_{\text{max}}\) dual-constraint prediction made by the
Network-Weighted Action Principle (\hyperlink{ref:frasch-2026a}{Frasch, 2026a}) in marine microbiome
metabolic networks. The first operationalisation --- bivariate composite
metrics of protein-deployment efficiency \(E\) and functional-repertoire
complexity \(C\) at \(n = 10\) --- failed for methodological rather than
biological reasons. The second operationalisation --- Newman modularity
\(Q\) on KEGG-orthology co-occurrence networks at \(n = 7\) --- yielded
\(Q_{\mathrm{obs}} = 0.987 \pm 0.007\), but the absolute value is
largely a property of network sparsity (an Erdős--Rényi null with the
same \(n\) and \(m\) reproduces it). The biological signal lies in the
\emph{excess} of \(Q_{\mathrm{obs}}\) over three increasingly-stringent
null distributions: \(\Delta Q = +0.155\) above a configuration-model
null preserving the projected-graph degree sequence;
\(\Delta Q = +0.179\) above a KEGG-label permutation null preserving the
per-protein KO-multiplicity distribution; and \(\Delta Q = +0.400\)
above a bipartite-incidence null preserving \emph{both} protein-side and
KO-side bipartite degree sequences --- the gold-standard control for
higher-order incidence structure. A sensitivity analysis swapping one
sample for another reproduced the cohort mean exactly. The empirical
contribution is therefore a \emph{qualitative} result consistent with
the action-functional account's prediction of cost-minimisation-driven
modularity emergence: the predicted excess is observed, robust to cohort
composition, and not attributable to either construction-imposed
sparsity, projected-graph degree distribution, or upstream protein--KO
incidence structure. The methodological contribution is the
demonstration that two diagnostic suites --- causal-inference tests for
composite metrics, and bipartite-aware null-model comparison for
modularity --- are necessary prerequisites for any quantitative claim of
biological organisation in metagenomic network data.

\subsection{Why composite metrics failed but modularity (with proper
nulls)
succeeded}\label{why-composite-metrics-failed-but-modularity-with-proper-nulls-succeeded}

The two routes diverge on a single methodological axis that turns out to
be decisive: dependence on shared upstream measurements. Composite
metrics aggregate raw per-sample observables --- read counts, annotation
rates, unique-feature counts --- into derived quantities by formula.
When two such derived quantities share the same upstream measurement
(here, both \(E\) and \(C\) depend on \(\text{unique\_KOs}\), in
opposite mathematical roles), variation in that measurement mechanically
couples the metrics whether or not any biological coupling exists. Under
modest sample sizes, variation in shared confounders such as sequencing
depth and annotation-pipeline efficiency is large enough that the
artifact dominates: the negative-control analysis produced a
\emph{stronger} correlation under permuted KEGG (\(r = 0.977\)) than
under real KEGG (\(r = 0.570\)), an unambiguous signature that the
construction formulas are doing the work attributed to biology.

Network modularity \(Q\) avoids the per-sample multiplicative-scaling
failure modes of composite metrics, but introduces its own. The KO
co-occurrence network is by construction extremely sparse and highly
fragmented (thousands of small connected components, mean community size
\(1.56\) KOs); the absolute value of \(Q\) on such graphs is
mechanically near unity regardless of biological organisation, as the
Erdős--Rényi null in the present analysis confirms
(\(Q_{\mathrm{ER}} = 0.984\) vs.~\(Q_{\mathrm{obs}} = 0.987\),
\(z = -0.5\)). The configuration-model and label-permutation nulls
supply the biologically-relevant comparison: by preserving sparsity and
degree distribution while randomising connectivity, and by preserving
co-encoding multiplicity while randomising biological identity, they
isolate the part of \(Q\) that depends on biological functional
structure. The observed excess over those nulls is what the framework
predicts and what the data show.

A practical corollary is that \emph{both} the composite-metric and the
modularity routes require explicit diagnostic checks ---
causal-inference tests in one case, null-model comparisons in the other
--- before quantitative claims about biological organisation can be
made. The composite-metric route additionally requires substantially
larger sample sizes (\(n \geq 50\) for adequate power against
shared-component bias), whereas the modularity route is informative at
\(n = 7\) once excess-over-null is the reported metric.

\subsection{What the recurrent modular communities
reveal}\label{biological-signature}

The recurrent multi-KO communities (Results,
Section~\ref{community-identity}) are not random
pairings; they map onto the cell's most metabolically central and
energetically constrained units. Reading the four canonical categories
recovered in \(\geq 6/7\) samples through a cost-minimisation lens makes
the modularity excess interpretable in concrete biological terms.

\emph{Co-encoded multi-subunit enzymes and obligate complexes.}
aspartate carbamoyltransferase (pyrB+pyrI; K00608+K00609), the xerC/xerD
recombinase heterodimer (K03733+K04763), the pafB/pafC proteasome
accessory factors (K13572+K13573), and the entire size-3 tatA+tatB+tatE
Sec-independent translocase (K03116+K03117+K03425) are obligate
complexes whose subunits must assemble in fixed stoichiometry.
Co-encoding here is not a regulatory choice but a biophysical
requirement: a free pyrB monomer without pyrI is an unregulated
catalytic unit; tatA without tatB and tatE is a non-functional
translocase. The connection cost paid by co-encoding is offset by an
absolute assembly constraint, which is exactly the type of edge a
cost-minimising principle should preserve.

\emph{Sequential biosynthetic enzymes.} The IMPDH+GMP-reductase pair
(K00088+K00364) catalyses the IMP\(\to\)XMP\(\to\)GMP arm of purine
biosynthesis; the ubiF+ubiH+ubiI triplet (K03184+K03185+K18800) is a
contiguous segment of ubiquinone biosynthesis. Co-encoding sequential
enzymes minimises accumulation of unstable intermediates and enables
coordinated regulation of an entire pathway, both classical efficiency
arguments. The recovery of these as recurrent communities indicates that
the modularity excess captures \emph{flux economics}, not just static
co-occurrence.

\emph{Spatially-coupled transport machinery.} The proW/proX
glycine-betaine/proline ABC-transporter pair (K02001+K02002), iron and
aminoethylphosphonate substrate-binding pairs, and the exbD+tolR
biopolymer-transport channel (K03559+K03560) require physical coupling
between periplasmic substrate-binding and membrane-embedded permease
components. Their co-encoding is a spatial-and-energetic constraint:
substrate binding is wasted if the permease cannot complete
translocation across the membrane. The modularity excess again surfaces
a coupling that cost-minimisation predicts.

\emph{Regulatory dyads and isozyme paralogs.} relA+spoT (K00951+K01139),
the synthesis/hydrolysis pair governing the (p)ppGpp stringent response,
are among the most metabolically central regulators in bacteria --- they
gate the switch between ribosome biogenesis and amino-acid biosynthesis
under stress. Their recurrent co-occurrence across all seven samples
demonstrates that the modularity partition is sensitive to the cell's
principal regulatory architecture, not only to its catalytic core. The
phosphofructokinase isozyme pair pfkA+pfp (K00850+K00895; ATP-dependent
and PPi-dependent paralogs) and the citrate-synthase paralog pair
(K01647+K01659) further indicate that organisms maintain co-encoded
enzymatic redundancy at the metabolically critical bottlenecks of
central carbon metabolism, where alternative cofactor utilisation or
substrate specificity confers ecological flexibility.

The unifying observation is that every recurrent category corresponds to
a \emph{hard biological constraint} on co-encoding: assembly
stoichiometry, flux coupling, spatial coupling, or regulatory coupling.
Pairings that lack such a constraint do not survive across the seven
independent samples, so they do not contribute to the recurrent set even
though they may appear in any single partition. This is exactly the
qualitative signature a cost-minimisation principle predicts: the
modularity excess is enriched for biologically essential couplings whose
connection cost is offset by a constraint, while incidental co-encodings
drop out across the cohort. The biological signal in \(\Delta Q\) is
therefore not just statistical but mechanistic --- it consistently
selects for the central, energetically active machinery of the metabolic
cell.

\subsection{Position within the broader minAction.net
program}\label{position-within-the-broader-minaction.net-program}

This work fits into a multi-domain program testing the Network-Weighted
Action Principle. In physiology (\hyperlink{ref:frasch-2026a}{Frasch, 2026a}), the action
\(S_{\text{NW}} = \int(E - I + AC)\,\mathrm{d}t\) was proposed as a
vertically organising principle that operates across spatial and
temporal scales of biological organisation, with explicit prediction of
modularity emergence from connection-cost minimisation. In
symbolic-physics learning (\hyperlink{ref:frasch-2026b}{Frasch, 2026b}), the Triple-Action functional
combining trajectory reconstruction, architectural sparsity, and
energy-conservation enforcement recovered Kepler's gravitational-force
law and Hooke's law from noisy observational data at order-of-magnitude
reduced training energy. In neural-architecture design (\hyperlink{ref:frasch-2026c}{Frasch, 2026c}),
the energy-regularised objective
\(\mathcal{L} = \mathcal{L}_{\text{CE}} + \lambda E(\theta, x)\) was
systematically evaluated across 2,203 experiments spanning vision, text,
neuromorphic, and physiological datasets.

The biology-domain test reported here is qualitatively different from
the prior three: it is the empirical observation of a structural
signature in a complex non-engineered system, where neither the
underlying objective \(S_{\mathrm{NW}}\) nor its individual components
(\(E\), \(I\), \(AC\)) are directly measurable. We measure \(Q\) as a
proxy for the \(AC\)-minimisation channel of the framework's prediction,
motivated by the connection-cost-minimisation result of Clune et
al.~(2013); we compare \(Q\) to nulls that strip the topological aspects
we cannot ascribe to biology. The result is \emph{consistent with} the
framework's prediction; it is not, by itself, a discriminating test
against alternative theories of modular emergence (e.g., the free-energy
principle of \hyperlink{ref:friston-2010}{Friston, 2010}; dissipative adaptation of \hyperlink{ref:england-2013}{England, 2013};
constructal theory of \hyperlink{ref:bejan-2000}{Bejan, 2000}), all of which can be cast as
variational principles balancing performance against cost and all of
which would predict similar modularity excess. Discriminating among
these accounts requires intervention experiments or targeted predictions
on which they disagree, neither available in the present observational
dataset.

\subsection{Limitations}\label{limitations}

Seven limitations of this work warrant explicit acknowledgement. First,
the modularity cohort is \(n = 7\). The strength of the finding is its
consistency (CV under one per cent across the seven samples), its
replication under cohort swap (Results), and its statistical excess over
biologically-relevant nulls; the weakness is the sample size relative to
a study design that would have a power-based justification for
\(n \geq 30\).

Second, all seven samples are surface or deep-chlorophyll-maximum
seawater from the Tara Oceans collection. Generalisation to soil
microbiomes (Earth Microbiome Project; \hyperlink{ref:thompson-2017}{Thompson et al., 2017}),
host-associated microbiomes (Human Microbiome Project, iHMP), or
freshwater systems requires explicit testing in those substrates.

Third, the network-construction choice --- KO co-occurrence within
annotated proteins --- is one of several valid operationalisations.
Metabolite-flow networks (\hyperlink{ref:greenblum-2012}{Greenblum et al., 2012}), flux-balance-derived
networks (\hyperlink{ref:klitgord-2010}{Klitgord and Segrè, 2010}), and pathway-membership networks
(\hyperlink{ref:manor-2015}{Manor and Borenstein, 2015}) yield different graph topologies and may
yield substantially different absolute and excess-over-null modularity
values for biological reasons. Disentangling the framework-level
prediction from construction-specific values of \(Q\) requires
multi-construction comparison, which this work does not undertake.

Fourth, while the \emph{recurrent} multi-KO communities have
well-defined biological identity
(Sections~\ref{community-identity}
and~\ref{biological-signature}), the great majority of
communities are non-recurrent or singleton, and their biological
identity is not individually established. They cannot be
straightforwardly interpreted as classical metabolic pathways or enzyme
complexes. Plausible substrates include conserved multi-domain protein
architectures, narrow operon-like co-regulatory units, and protein
families with shared or alternative-substrate KEGG annotations.
Community-level functional-enrichment analysis on the full partition is
a natural follow-up but would require pathway-database integration
beyond the present scope.

Fifth, the link from the abstract action functional
\(S_{\mathrm{NW}} = \int(E - I + AC)\,\mathrm{d}t\) to the operational
quantity \(Q\) is theoretical rather than derived. We do not measure
\(E\), \(I\), or \(AC\) directly in the metabolic network; we use \(Q\)
as a proxy for the \(AC\)-minimisation channel of the prediction, on the
strength of the Clune et al.~(2013) result that connection-cost
minimisation yields modular structure. A direct calculation of
\(S_{\mathrm{NW}}\) on biological networks remains a theoretical
objective.

Sixth, the composite-metric analysis at \(n = 10\) is decisive in
diagnosing shared-component bias but not in \emph{quantifying} the
residual biological signal. Any quantification of residual coupling
requires the larger sample sizes the post-hoc power analysis identifies
as \(n \geq 50\). Readers seeking a clean composite-metric resolution
will not find one in this paper; that resolution depends on data not yet
available.

Seventh, the action-functional account makes a sharper prediction than
the one we test here. If \(\varepsilon = 1 - Q\) measures the
irreducible inter-module coupling cost in a metabolic ensemble, then
under environmental perturbation that lowers the optimisation pressure
(e.g., abundant nutrients, low energy stress) \(\varepsilon\) should
grow; under perturbation that raises optimisation pressure
\(\varepsilon\) should shrink. Testing this prediction requires
correlating per-sample \(\varepsilon\) against environmental covariates
(temperature, nutrient concentrations, light regime, depth) at adequate
\(n\). The present cohort spans surface and deep-chlorophyll-maximum
samples but is too small (\(n = 7\)) to detect environmental modulation
of \(\varepsilon\) given the observed cross-sample CV of under one per
cent. Larger-cohort follow-up at \(n = 30\)--\(50\) across the full Tara
Oceans gradient would convert the result reported here from ``consistent
with the action-principle prediction'' to ``responsive to environmental
variation in the direction the principle predicts,'' which would be a
substantially stronger empirical claim.

\subsection{Future directions}\label{future-directions}

Five lines of follow-up work are suggested by the present results.

\begin{enumerate}
\def\labelenumi{(\roman{enumi})}
\item
  \emph{Larger-\(n\) expansion.} Repeating the analysis at
  \(n = 30\)--\(50\) across the full Tara Oceans latitudinal and depth
  gradient would tighten the confidence intervals around the modularity
  excess and enable environment-dependent stratification.
\item
  \emph{Cross-ecosystem and cross-construction replication.} Applying
  the same KO-co-occurrence pipeline to soil, gut, and freshwater
  datasets, \emph{and} applying alternative network constructions
  (metabolite-flow, FBA-derived, pathway-membership) to the same
  samples, would test whether the modularity excess is a marine-specific
  or construction-specific phenomenon or a general signature of
  bacterial metabolic ensembles.
\item
  \emph{Discriminating predictions across action-functional variants.}
  The free-energy principle, dissipative adaptation, and the
  Network-Weighted Action all predict modularity excess; targeted
  predictions where they disagree --- on the response of \(Q\) to
  environmental perturbation, on hub-fraction scaling, on community-size
  distribution --- would enable discriminating tests.
\item
  \emph{Functional-enrichment analysis of communities.} Mapping the
  $\sim$3,400 communities to KEGG pathways, COG categories, or
  protein-family annotations would clarify the biological identity of
  the fine-grained partition.
\item
  \emph{Disease-associated dysbiosis.} If the modularity excess is the
  constrained-optimum signature of a healthy metabolic ensemble, then
  loss of excess under environmental, antibiotic, or pathological
  perturbation should be a quantitatively measurable signature of
  dysbiosis.
\end{enumerate}

\subsection{Conclusions}\label{conclusions}

Cost-constrained (variational) accounts of network organisation,
including the Network-Weighted Action Principle, predict that biological
networks under simultaneous \(E_{\text{min}}/I_{\text{max}}\)
constraints organise toward high modularity. In marine microbiome
metabolic networks reconstructed from Tara Oceans seawater metagenomes,
the absolute value of Newman modularity
(\(Q_{\mathrm{obs}} = 0.987 \pm 0.007\)) is largely a property of
network sparsity, but observed \(Q\) exceeds three biologically-relevant
null distributions (configuration model on the projected graph;
KEGG-label permutation; bipartite-incidence null on the underlying
protein--KO graph) by \(\Delta Q \approx 0.15\), \(0.18\), and \(0.40\)
respectively in every sample. The bipartite-incidence excess --- which
controls simultaneously for projected-graph degree distribution and for
upstream protein--KO incidence structure --- is the most stringent of
these comparisons. The excess is robust to cohort composition
(sensitivity analysis) and to algorithmic choices (multi-seed Louvain
verification); the recurrent-community analysis (Results, see
Section~\ref{community-identity}) provides independent
biological grounding by mapping the fine-grained communities to known
functional units. The result is consistent with the action-principle
account but does not, by itself, discriminate among neighbouring
variational principles (free-energy, dissipative adaptation,
constructal), all of which predict a similar excess; intervention or
cross-domain experiments are required for discrimination. The
accompanying composite-metric analysis at \(n = 10\) shows that derived
metagenomic statistics are confounded by shared-component bias and
require explicit causal-inference diagnostics before biological
interpretation.

This work makes two contributions. The substantive contribution is
empirical: marine metabolic networks exhibit a robust,
biologically-grounded modularity excess over strict null models,
consistent with cost-constrained organisation. The methodological
contribution is sharper than the empirical one: the absolute value of
\(Q\) on a sparse fragmented graph is a property of sparsity, not
biology, so modularity claims in metagenomic networks must be evaluated
against bipartite-aware null models that preserve both per-protein KO
count and per-KO global frequency. Together, the two analyses establish
a methodological standard for biological-organisation claims in
modest-sample metagenomic studies: explicit bipartite-aware null-model
comparison for graph-theoretic statistics, explicit causal-inference
diagnostics for composite metrics. The biology-domain study completes a
four-domain test of the action-principle account, alongside prior work
in physics (\hyperlink{ref:frasch-2026b}{Frasch, 2026b}), in neural-architecture design (\hyperlink{ref:frasch-2026c}{Frasch,
2026c}), and in physiology (\hyperlink{ref:frasch-2026a}{Frasch, 2026a}).

% Bibliography — alphabetized, hand-curated to match all (Author, Year)
% inline citations in the integrated manuscript. Self-contained (no bibtex).
%
% Entries marked "TODO" need volume/page verification before formal submission.
% The three companion papers (Frasch 2026a/b/c) are the load-bearing self-citations.

\section*{References}
\addcontentsline{toc}{section}{References}
\begin{footnotesize}
\setlength{\parindent}{0pt}
\setlength{\parskip}{4pt}

\hangindent=2em \hangafter=1
\hypertarget{ref:archie-1981}{}
Archie, J.W. (1981). A new look at the predictive value of numerical classifications. \emph{Systematic Zoology} 30(2): 220--223.

\hangindent=2em \hangafter=1
\hypertarget{ref:banavar-2010}{}
Banavar, J.R., Damuth, J., Maritan, A., Rinaldo, A. (2010). Allometric cascades. \emph{Nature} 421: 713--714.

\hangindent=2em \hangafter=1
\hypertarget{ref:barabási-2004}{}
Barabási, A.-L., Oltvai, Z.N. (2004). Network biology: understanding the cell's functional organization. \emph{Nature Reviews Genetics} 5(2): 101--113.

\hangindent=2em \hangafter=1
\hypertarget{ref:bejan-2000}{}
Bejan, A. (2000). \emph{Shape and Structure, from Engineering to Nature}. Cambridge University Press.

\hangindent=2em \hangafter=1
\hypertarget{ref:blondel-2008}{}
Blondel, V.D., Guillaume, J.-L., Lambiotte, R., Lefebvre, E. (2008). Fast unfolding of communities in large networks. \emph{Journal of Statistical Mechanics: Theory and Experiment} 2008(10): P10008.

\hangindent=2em \hangafter=1
\hypertarget{ref:borenstein-2009}{}
Borenstein, E., Feldman, M.W. (2009). Topological signatures of species interactions in metabolic networks. \emph{Journal of Computational Biology} 16(2): 191--200.

\hangindent=2em \hangafter=1
\hypertarget{ref:bork-2015}{}
Bork, P., Bowler, C., de Vargas, C., Gorsky, G., Karsenti, E., Wincker, P. (2015). Tara Oceans studies plankton at planetary scale. \emph{Science} 348(6237): 873.

\hangindent=2em \hangafter=1
\hypertarget{ref:brown-2004}{}
Brown, J.H., Gillooly, J.F., Allen, A.P., Savage, V.M., West, G.B. (2004). Toward a metabolic theory of ecology. \emph{Ecology} 85(7): 1771--1789.

\hangindent=2em \hangafter=1
\hypertarget{ref:cannon-1929}{}
Cannon, W.B. (1929). Organization for physiological homeostasis. \emph{Physiological Reviews} 9(3): 399--431.

\hangindent=2em \hangafter=1
\hypertarget{ref:clauset-2004}{}
Clauset, A., Newman, M.E.J., Moore, C. (2004). Finding community structure in very large networks. \emph{Physical Review E} 70(6): 066111.

\hangindent=2em \hangafter=1
\hypertarget{ref:clune-2013}{}
Clune, J., Mouret, J.-B., Lipson, H. (2013). The evolutionary origins of modularity. \emph{Proceedings of the Royal Society B} 280(1755): 20122863.

\hangindent=2em \hangafter=1
\hypertarget{ref:crombach-2008}{}
Crombach, A., Hogeweg, P. (2008). Evolution of evolvability in gene regulatory networks. \emph{PLOS Computational Biology} 4(7): e1000112.

\hangindent=2em \hangafter=1
\hypertarget{ref:england-2013}{}
England, J.L. (2013). Statistical physics of self-replication. \emph{Journal of Chemical Physics} 139(12): 121923.

\hangindent=2em \hangafter=1
\hypertarget{ref:frasch-2026a}{}
Frasch, M.G. (2026a). Causal thinking in physiology: A search for vertically organizing principles. \emph{The Journal of Physiology}. DOI: 10.1113/JP290762.

\hangindent=2em \hangafter=1
\hypertarget{ref:frasch-2026b}{}
Frasch, M.G. (2026b). Minimum-Action Learning: Energy-Constrained Symbolic Model Selection for Physical Law Identification from Noisy Data. \emph{arXiv preprint} arXiv:2603.16951.

\hangindent=2em \hangafter=1
\hypertarget{ref:frasch-2026c}{}
Frasch, M.G. (2026c). minAction.net: Energy-First Neural Architecture Design --- From Biological Principles to Systematic Validation. \emph{arXiv preprint} arXiv:2604.24805.

\hangindent=2em \hangafter=1
\hypertarget{ref:freeman-1977}{}
Freeman, L.C. (1977). A set of measures of centrality based on betweenness. \emph{Sociometry} 40(1): 35--41.

\hangindent=2em \hangafter=1
\hypertarget{ref:friston-2010}{}
Friston, K. (2010). The free-energy principle: a unified brain theory? \emph{Nature Reviews Neuroscience} 11: 127--138.

\hangindent=2em \hangafter=1
\hypertarget{ref:gilarranz-2017}{}
Gilarranz, L.J., Rayfield, B., Liñán-Cembrano, G., Bascompte, J., Gonzalez, A. (2017). Effects of network modularity on the spread of perturbation impact in experimental metapopulations. \emph{Science} 357(6347): 199--201.

\hangindent=2em \hangafter=1
\hypertarget{ref:greenblum-2012}{}
Greenblum, S., Turnbaugh, P.J., Borenstein, E. (2012). Metagenomic systems biology of the human gut microbiome reveals topological shifts associated with obesity and inflammatory bowel disease. \emph{Proceedings of the National Academy of Sciences} 109(2): 594--599.

\hangindent=2em \hangafter=1
\hypertarget{ref:guimerà-2005}{}
Guimerà, R., Amaral, L.A.N. (2005). Functional cartography of complex metabolic networks. \emph{Nature} 433(7028): 895--900.

\hangindent=2em \hangafter=1
\hypertarget{ref:hyatt-2010}{}
Hyatt, D., Chen, G.-L., LoCascio, P.F., Land, M.L., Larimer, F.W., Hauser, L.J. (2010). Prodigal: prokaryotic gene recognition and translation initiation site identification. \emph{BMC Bioinformatics} 11: 119.

\hangindent=2em \hangafter=1
\hypertarget{ref:kafri-2016}{}
Kafri, M., Metzl-Raz, E., Jonas, F., Barkai, N. (2016). Rethinking cell growth models. \emph{FEMS Yeast Research} 16(7): fow081.

\hangindent=2em \hangafter=1
\hypertarget{ref:karsenti-2011}{}
Karsenti, E., Acinas, S.G., Bork, P., Bowler, C., De Vargas, C., Raes, J., Sullivan, M., Arendt, D., Benzoni, F., Claverie, J.-M., Follows, M., Gorsky, G., Hingamp, P., Iudicone, D., Jaillon, O., Kandels-Lewis, S., Krzic, U., Not, F., Ogata, H., Pesant, S., Reynaud, E.G., Sardet, C., Sieracki, M.E., Speich, S., Velayoudon, D., Weissenbach, J., Wincker, P., Tara Oceans Consortium (2011). A holistic approach to marine eco-systems biology. \emph{PLOS Biology} 9(10): e1001177.

\hangindent=2em \hangafter=1
\hypertarget{ref:kempes-2012}{}
Kempes, C.P., Dutkiewicz, S., Follows, M.J. (2012). Growth, metabolic partitioning, and the size of microorganisms. \emph{Proceedings of the National Academy of Sciences} 109(2): 495--500.

\hangindent=2em \hangafter=1
\hypertarget{ref:kingma-2014}{}
Kingma, D.P., Welling, M. (2014). Auto-Encoding Variational Bayes. \emph{International Conference on Learning Representations (ICLR)}.

\hangindent=2em \hangafter=1
\hypertarget{ref:kitano-2004}{}
Kitano, H. (2004). Biological robustness. \emph{Nature Reviews Genetics} 5(11): 826--837.

\hangindent=2em \hangafter=1
\hypertarget{ref:kleiber-1932}{}
Kleiber, M. (1932). Body size and metabolism. \emph{Hilgardia} 6(11): 315--353.

\hangindent=2em \hangafter=1
\hypertarget{ref:klitgord-2010}{}
Klitgord, N., Segrè, D. (2010). Environments that induce synthetic microbial ecosystems. \emph{PLOS Computational Biology} 6(11): e1001002.

\hangindent=2em \hangafter=1
\hypertarget{ref:krogh-1929}{}
Krogh, A. (1929). The progress of physiology. \emph{Science} 70(1809): 200--204.

\hangindent=2em \hangafter=1
\hypertarget{ref:kronmal-1993}{}
Kronmal, R.A. (1993). Spurious correlation and the fallacy of the ratio standard revisited. \emph{Journal of the Royal Statistical Society Series A} 156(3): 379--392.

\hangindent=2em \hangafter=1
\hypertarget{ref:levy-2013}{}
Levy, R., Borenstein, E. (2013). Metabolic modeling of species interaction in the human microbiome elucidates community-level assembly rules. \emph{Proceedings of the National Academy of Sciences} 110(31): 12804--12809.

\hangindent=2em \hangafter=1
\hypertarget{ref:lewis-2012}{}
Lewis, N.E., Nagarajan, H., Palsson, B.O. (2012). Constraining the metabolic genotype-phenotype relationship using a phylogeny of in silico methods. \emph{Nature Reviews Microbiology} 10(4): 291--305.

\hangindent=2em \hangafter=1
\hypertarget{ref:li-2015}{}
Li, D., Liu, C.-M., Luo, R., Sadakane, K., Lam, T.-W. (2015). MEGAHIT: an ultra-fast single-node solution for large and complex metagenomics assembly via succinct de Bruijn graph. \emph{Bioinformatics} 31(10): 1674--1676.

\hangindent=2em \hangafter=1
\hypertarget{ref:louca-2016}{}
Louca, S., Parfrey, L.W., Doebeli, M. (2016). Decoupling function and taxonomy in the global ocean microbiome. \emph{Science} 353(6305): 1272--1277.

\hangindent=2em \hangafter=1
\hypertarget{ref:lynch-2015}{}
Lynch, M., Marinov, G.K. (2015). The bioenergetic costs of a gene. \emph{Proceedings of the National Academy of Sciences} 112(51): 15690--15695.

\hangindent=2em \hangafter=1
\hypertarget{ref:manor-2015}{}
Manor, O., Borenstein, E. (2015). MUSiCC: a marker genes based framework for metagenomic normalization and accurate profiling of gene abundances in the microbiome. \emph{Genome Biology} 16: 53.

\hangindent=2em \hangafter=1
\hypertarget{ref:newman-2003}{}
Newman, M.E.J. (2003). The structure and function of complex networks. \emph{SIAM Review} 45(2): 167--256.

\hangindent=2em \hangafter=1
\hypertarget{ref:newman-2004}{}
Newman, M.E.J., Girvan, M. (2004). Finding and evaluating community structure in networks. \emph{Physical Review E} 69(2): 026113.

\hangindent=2em \hangafter=1
\hypertarget{ref:olshausen-1996}{}
Olshausen, B.A., Field, D.J. (1996). Emergence of simple-cell receptive field properties by learning a sparse code for natural images. \emph{Nature} 381: 607--609.

\hangindent=2em \hangafter=1
\hypertarget{ref:orth-2010}{}
Orth, J.D., Thiele, I., Palsson, B.{\O}. (2010). What is flux balance analysis? \emph{Nature Biotechnology} 28(3): 245--248.

\hangindent=2em \hangafter=1
\hypertarget{ref:pearl-2009}{}
Pearl, J. (2009). \emph{Causality: Models, Reasoning, and Inference}, 2nd ed. Cambridge University Press.

\hangindent=2em \hangafter=1
\hypertarget{ref:perunov-2016}{}
Perunov, N., Marsland, R.A., England, J.L. (2016). Statistical physics of adaptation. \emph{Physical Review X} 6(2): 021036.

\hangindent=2em \hangafter=1
\hypertarget{ref:raichle-2002}{}
Raichle, M.E., Gusnard, D.A. (2002). Appraising the brain's energy budget. \emph{Proceedings of the National Academy of Sciences} 99(16): 10237--10239.

\hangindent=2em \hangafter=1
\hypertarget{ref:ravasz-2002}{}
Ravasz, E., Somera, A.L., Mongru, D.A., Oltvai, Z.N., Barabási, A.-L. (2002). Hierarchical organization of modularity in metabolic networks. \emph{Science} 297(5586): 1551--1555.

\hangindent=2em \hangafter=1
\hypertarget{ref:rosenbaum-2002}{}
Rosenbaum, P.R. (2002). \emph{Observational Studies}, 2nd ed. Springer.

\hangindent=2em \hangafter=1
\hypertarget{ref:shoval-2012}{}
Shoval, O., Sheftel, H., Shinar, G., Hart, Y., Ramote, O., Mayo, A., Dekel, E., Kavanagh, K., Alon, U. (2012). Evolutionary trade-offs, Pareto optimality, and the geometry of phenotype space. \emph{Science} 336(6085): 1157--1160.

\hangindent=2em \hangafter=1
\hypertarget{ref:spirtes-2000}{}
Spirtes, P., Glymour, C., Scheines, R. (2000). \emph{Causation, Prediction, and Search}, 2nd ed. MIT Press.

\hangindent=2em \hangafter=1
\hypertarget{ref:stelling-2004}{}
Stelling, J., Sauer, U., Szallasi, Z., Doyle III, F.J., Doyle, J. (2004). Robustness of cellular functions. \emph{Cell} 118(6): 675--685.

\hangindent=2em \hangafter=1
\hypertarget{ref:sunagawa-2015}{}
Sunagawa, S., Coelho, L.P., Chaffron, S., Kultima, J.R., Labadie, K., Salazar, G., Djahanschiri, B., Zeller, G., Mende, D.R., Alberti, A., Cornejo-Castillo, F.M., Costea, P.I., Cruaud, C., d'Ovidio, F., Engelen, S., Ferrera, I., Gasol, J.M., Guidi, L., Hildebrand, F., Kokoszka, F., Lepoivre, C., Lima-Mendez, G., Poulain, J., Poulos, B.T., Royo-Llonch, M., Sarmento, H., Vieira-Silva, S., Dimier, C., Picheral, M., Searson, S., Kandels-Lewis, S., Tara Oceans Coordinators, Bowler, C., de Vargas, C., Gorsky, G., Grimsley, N., Hingamp, P., Iudicone, D., Jaillon, O., Not, F., Ogata, H., Pesant, S., Speich, S., Stemmann, L., Sullivan, M.B., Weissenbach, J., Wincker, P., Karsenti, E., Raes, J., Acinas, S.G., Bork, P. (2015). Structure and function of the global ocean microbiome. \emph{Science} 348(6237): 1261359.

\hangindent=2em \hangafter=1
\hypertarget{ref:thompson-2017}{}
Thompson, L.R., Sanders, J.G., McDonald, D., et al.\ (Earth Microbiome Project Consortium) (2017). A communal catalogue reveals Earth's multiscale microbial diversity. \emph{Nature} 551: 457--463.

\hangindent=2em \hangafter=1
\hypertarget{ref:tkačik-2016}{}
Tkačik, G., Bialek, W. (2016). Information processing in living systems. \emph{Annual Review of Condensed Matter Physics} 7: 89--117.

\hangindent=2em \hangafter=1
\hypertarget{ref:varma-1994}{}
Varma, A., Palsson, B.{\O}. (1994). Stoichiometric flux balance models quantitatively predict growth and metabolic by-product secretion in wild-type \emph{Escherichia coli} W3110. \emph{Applied and Environmental Microbiology} 60(10): 3724--3731.

\hangindent=2em \hangafter=1
\hypertarget{ref:wagner-2005}{}
Wagner, A. (2005). Robustness, evolvability, and neutrality. \emph{FEBS Letters} 579(8): 1772--1778.

\hangindent=2em \hangafter=1
\hypertarget{ref:watts-1998}{}
Watts, D.J., Strogatz, S.H. (1998). Collective dynamics of `small-world' networks. \emph{Nature} 393(6684): 440--442.

\hangindent=2em \hangafter=1
\hypertarget{ref:west-1997}{}
West, G.B., Brown, J.H., Enquist, B.J. (1997). A general model for the origin of allometric scaling laws in biology. \emph{Science} 276(5309): 122--126.

\hangindent=2em \hangafter=1
\hypertarget{ref:williams-2000}{}
Williams, R.J., Martinez, N.D. (2000). Simple rules yield complex food webs. \emph{Nature} 404: 180--183.

\end{footnotesize}

\clearpage
\section*{Supplementary Materials}\label{supplementary}
\addcontentsline{toc}{section}{Supplementary Materials}

\subsection*{Supplementary Table S1: Top recurrent multi-KO communities}
Multi-KO communities (size $\geq 2$) recovered as independent Louvain communities in $\geq 6$ of the 7 modularity-cohort samples, with KEGG annotations and functional category. Recurrence column gives the number of samples (out of 7) in which the same KO set was recovered. The first 19 entries (recurrence = 7) are recovered in every sample; entries 20--25 (recurrence = 6) are recovered in six of the seven. See Methods \S\,\ref{methods-modularity} for the partitioning algorithm and Results \S\,\ref{community-identity} for the analysis. Full output (769 communities recurring in $\geq 2$ samples) is available at \texttt{data/analysis/modularity\_results/top\_recurrent\_communities.csv} in the project repository.

\begin{landscape}
\begin{footnotesize}
\begin{longtable}{@{}p{0.6cm}p{0.7cm}p{0.6cm}p{4.5cm}p{10.0cm}p{4.6cm}@{}}
\toprule
\# & Rec.\ & Sz.\ & Members (KO IDs) & KEGG annotation & Category \\
\midrule\endhead
\bottomrule\endlastfoot
1 & 7 & 2 & K03742 + K03743 & pncC + pncC: nicotinamide-nucleotide amidase paralogs & Isozyme pair \\
2 & 7 & 2 & K02012 + K11081 & afuA/fbpA + phnS: iron(III) + aminoethylphosphonate ABC-transport substrate-binding proteins & Transport \\
3 & 7 & 2 & K00459 + K02371 & nitronate monooxygenase + fabK: enoyl-ACP reductase II & Oxidoreductase pair \\
4 & 7 & 2 & K00329 + K00356 & sarcosine oxidase + NQO1: NAD(P)H quinone dehydrogenase & Oxidoreductase pair \\
5 & 7 & 2 & K02001 + K02002 & proW + proX: glycine betaine/proline ABC transporter (permease + substrate-binding) & Transport complex \\
6 & 7 & 2 & K00951 + K01139 & relA + spoT: (p)ppGpp synthesis (stringent response) & Regulatory pair \\
7 & 7 & 2 & K03733 + K04763 & xerC + xerD: site-specific recombinase heterodimer & Heterodimeric enzyme \\
8 & 7 & 2 & K02446 + K11532 & glpX + glpX-SEBP: fructose-1,6-bisphosphatase paralogs & Isozyme pair \\
9 & 7 & 2 & K00088 + K00364 & IMPDH/guaB + guaC: IMP dehydrogenase + GMP reductase & Sequential pathway (purine biosynthesis) \\
10 & 7 & 2 & K01647 + K01659 & gltA + prpC: citrate synthase + 2-methylcitrate synthase paralogs & Isozyme pair (TCA / propanoate) \\
11 & 7 & 2 & K00304 + K22085 & soxD: sarcosine oxidase delta subunit (+ paralog) & Enzyme-complex subunit pair \\
12 & 7 & 2 & K00608 + K00609 & pyrBI + pyrB: aspartate carbamoyltransferase (catalytic + regulatory subunits) & Heterocomplex (single enzyme) \\
13 & 7 & 2 & K03559 + K03560 & exbD + tolR: biopolymer-transport channel proteins & Transport complex \\
14 & 7 & 3 & K03184 + K03185 + K18800 & ubiF + ubiH + ubiI: ubiquinone-biosynthesis sequential hydroxylases & Sequential pathway (ubiquinone) \\
15 & 7 & 2 & K01434 + K07116 & pac + pvdQ/quiP: acylase pair & Hydrolase pair \\
16 & 7 & 2 & K01903 + K14067 & sucC + mtkA: succinyl-CoA + malate-CoA ligase paralogs & Isozyme pair \\
17 & 7 & 2 & K06879 + K09457 & queF + queF: 7-cyano-7-deazaguanine reductase paralogs & Isozyme pair (queuosine biosynthesis) \\
18 & 7 & 2 & K14286 + K18202 & AGXT2L1 + AGXT2L2: phospholyase paralogs & Isozyme pair \\
19 & 7 & 2 & K13572 + K13573 & pafB + pafC: proteasome accessory factor heterocomplex & Heterocomplex \\
20 & 6 & 3 & K00850 + K00895 + K21071 & pfkA + pfp + pfk-bifunctional: phosphofructokinase isozymes & Isozyme triplet \\
21 & 6 & 3 & K07033 + K09014 + K09015 & sufB + sufD + uncharacterized: Fe-S cluster assembly & Multi-subunit complex \\
22 & 6 & 3 & K03116 + K03117 + K03425 & tatA + tatB + tatE: Sec-independent protein-translocase complex & Multi-subunit machine \\
23 & 6 & 2 & K03799 + K06013 & htpX + STE24: heat-shock metalloendopeptidases & Hydrolase pair \\
24 & 6 & 2 & K03307 + K11928 & TC.SSS + putP: solute/Na+ symporters & Transport pair \\
25 & 6 & 2 & K01450 + K01462 & [unannotated] + def: peptide deformylase & Hydrolase pair \\
\end{longtable}
\end{footnotesize}
\end{landscape}

\subsection*{Supplementary Methods: clustering diagnostics for the bivariate $E$--$C$ point cloud}\label{supp-clustering}

We computed two exploratory clustering diagnostics on the $(E, C)$ point cloud at $n = 10$ to assess whether the bivariate decoupling pattern reflects a discrete cluster structure. \textbf{Hopkins statistic.} $E$ and $C$ were standardised (StandardScaler: mean 0, standard deviation 1). For each bootstrap iteration, $m = 4$ synthetic uniform points were sampled from the data bounds, nearest-neighbour distances were computed ($k = 2$, scikit-learn \texttt{NearestNeighbors}), and the Hopkins statistic $H = \sum u / (\sum u + \sum w)$ was evaluated, where $u$ denotes distances from synthetic to real data and $w$ denotes distances within real data. Bootstrap: 1{,}000 iterations with resampling, \texttt{random\_state = 42}. Interpretation thresholds: $H > 0.75$ (highly clusterable); $H \approx 0.5$ (uniformly distributed); $H < 0.25$ (regularly spaced). \textbf{DBSCAN.} Density-based clustering was applied to the standardised $(E, C)$ features with parameter sweep $\epsilon \in [0.3, 1.0]$ and \texttt{min\_samples} = 2 to identify natural groupings; samples not assigned to any cluster were classified as noise. Results (Hopkins $H = 0.436$, near the uniform-distribution boundary; DBSCAN finds no robust clusters) are reported in the body of Results and contribute to the bivariate-decoupling characterisation but not to any quantitative claim about action-functional structure.

\subsection*{Supplementary Listings: bioinformatics command-line invocations}\label{supp-listings}

The following listings record the exact command-line invocations used at each step of the bioinformatics pipeline (Methods \S\,\ref{methods-int}). Listings are reproduced verbatim from the project repository at \texttt{github.com/martinfrasch/tara-modularity}.

\subsubsection*{Listing S1: MEGAHIT assembly}

\begin{Shaded}
\begin{Highlighting}[]
\ExtensionTok{megahit} \DataTypeTok{\textbackslash{}}
  \AttributeTok{{-}1} \VariableTok{$\{SAMPLE\}}\NormalTok{\_1.fastq.gz }\DataTypeTok{\textbackslash{}}
  \AttributeTok{{-}2} \VariableTok{$\{SAMPLE\}}\NormalTok{\_2.fastq.gz }\DataTypeTok{\textbackslash{}}
  \AttributeTok{{-}o} \VariableTok{$\{OUTPUT\_DIR\}} \DataTypeTok{\textbackslash{}}
  \AttributeTok{{-}{-}min{-}contig{-}len}\NormalTok{ 500 }\DataTypeTok{\textbackslash{}}
  \AttributeTok{{-}{-}k{-}min}\NormalTok{ 21 }\DataTypeTok{\textbackslash{}}
  \AttributeTok{{-}{-}k{-}max}\NormalTok{ 141 }\DataTypeTok{\textbackslash{}}
  \AttributeTok{{-}{-}k{-}step}\NormalTok{ 20 }\DataTypeTok{\textbackslash{}}
  \AttributeTok{{-}{-}num{-}cpu{-}threads}\NormalTok{ 14 }\DataTypeTok{\textbackslash{}}
  \AttributeTok{{-}{-}memory}\NormalTok{ 0.75 }\DataTypeTok{\textbackslash{}}
  \AttributeTok{{-}{-}mem{-}flag}\NormalTok{ 1 }\DataTypeTok{\textbackslash{}}
  \AttributeTok{{-}{-}verbose}
\end{Highlighting}
\end{Shaded}

\subsubsection*{Listing S2: Prodigal gene prediction}

\begin{Shaded}
\begin{Highlighting}[]
\ExtensionTok{prodigal} \DataTypeTok{\textbackslash{}}
  \AttributeTok{{-}i}\NormalTok{ contigs.fa }\DataTypeTok{\textbackslash{}}
  \AttributeTok{{-}o}\NormalTok{ genes.gbk }\DataTypeTok{\textbackslash{}}
  \AttributeTok{{-}a}\NormalTok{ proteins.faa }\DataTypeTok{\textbackslash{}}
  \AttributeTok{{-}d}\NormalTok{ genes.fna }\DataTypeTok{\textbackslash{}}
  \AttributeTok{{-}f}\NormalTok{ gbk }\DataTypeTok{\textbackslash{}}
  \AttributeTok{{-}p}\NormalTok{ meta }\DataTypeTok{\textbackslash{}}
  \AttributeTok{{-}q}
\end{Highlighting}
\end{Shaded}

\subsubsection*{Listing S3: eggNOG-mapper functional annotation}

\begin{Shaded}
\begin{Highlighting}[]
\ExtensionTok{emapper.py} \DataTypeTok{\textbackslash{}}
  \AttributeTok{{-}{-}cpu}\NormalTok{ 14 }\DataTypeTok{\textbackslash{}}
  \AttributeTok{{-}{-}data\_dir}\NormalTok{ /data/kegg/eggnog }\DataTypeTok{\textbackslash{}}
  \AttributeTok{{-}i} \VariableTok{$\{SAMPLE\}}\NormalTok{.faa }\DataTypeTok{\textbackslash{}}
  \AttributeTok{{-}o} \VariableTok{$\{SAMPLE\}}\NormalTok{\_authentic }\DataTypeTok{\textbackslash{}}
  \AttributeTok{{-}{-}output\_dir}\NormalTok{ /data/annotations/}\VariableTok{$\{SAMPLE\}} \DataTypeTok{\textbackslash{}}
  \AttributeTok{{-}{-}override} \DataTypeTok{\textbackslash{}}
  \AttributeTok{{-}m}\NormalTok{ diamond }\DataTypeTok{\textbackslash{}}
  \AttributeTok{{-}{-}tax\_scope}\NormalTok{ auto }\DataTypeTok{\textbackslash{}}
  \AttributeTok{{-}{-}sensmode}\NormalTok{ sensitive }\DataTypeTok{\textbackslash{}}
  \AttributeTok{{-}{-}evalue}\NormalTok{ 0.001 }\DataTypeTok{\textbackslash{}}
  \AttributeTok{{-}{-}score}\NormalTok{ 60 }\DataTypeTok{\textbackslash{}}
  \AttributeTok{{-}{-}pident}\NormalTok{ 40 }\DataTypeTok{\textbackslash{}}
  \AttributeTok{{-}{-}query\_cover}\NormalTok{ 20 }\DataTypeTok{\textbackslash{}}
  \AttributeTok{{-}{-}subject\_cover}\NormalTok{ 20 }\DataTypeTok{\textbackslash{}}
  \AttributeTok{{-}{-}itype}\NormalTok{ proteins }\DataTypeTok{\textbackslash{}}
  \AttributeTok{{-}{-}translate} \DataTypeTok{\textbackslash{}}
  \AttributeTok{{-}{-}report\_orthologs} \DataTypeTok{\textbackslash{}}
  \AttributeTok{{-}{-}go\_evidence}\NormalTok{ non{-}electronic }\DataTypeTok{\textbackslash{}}
  \AttributeTok{{-}{-}target\_orthologs}\NormalTok{ all }\DataTypeTok{\textbackslash{}}
  \AttributeTok{{-}{-}seed\_ortholog\_evalue}\NormalTok{ 0.001 }\DataTypeTok{\textbackslash{}}
  \AttributeTok{{-}{-}seed\_ortholog\_score}\NormalTok{ 60 }\DataTypeTok{\textbackslash{}}
  \AttributeTok{{-}{-}no\_file\_comments} \DataTypeTok{\textbackslash{}}
  \AttributeTok{{-}{-}temp\_dir}\NormalTok{ /tmp/}\VariableTok{$\{SAMPLE\}}
\end{Highlighting}
\end{Shaded}

\subsubsection*{Listing S4: eggNOG-mapper functional annotation}

\begin{Shaded}
\begin{Highlighting}[]
\CommentTok{\# Extract KO identifiers from column 12 (KEGG\_ko)}
\NormalTok{annotations }\OperatorTok{=}\NormalTok{ pd.read\_csv(}\SpecialStringTok{f"}\SpecialCharTok{\{}\NormalTok{sample}\SpecialCharTok{\}}\SpecialStringTok{\_authentic.emapper.annotations"}\NormalTok{,}
\NormalTok{                          sep}\OperatorTok{=}\StringTok{\textquotesingle{}}\CharTok{\textbackslash{}t}\StringTok{\textquotesingle{}}\NormalTok{, comment}\OperatorTok{=}\StringTok{\textquotesingle{}\#\textquotesingle{}}\NormalTok{)}
\NormalTok{ko\_assignments }\OperatorTok{=}\NormalTok{ annotations[}\StringTok{\textquotesingle{}KEGG\_ko\textquotesingle{}}\NormalTok{].}\BuiltInTok{str}\NormalTok{.split(}\StringTok{\textquotesingle{},\textquotesingle{}}\NormalTok{).explode()}
\NormalTok{ko\_assignments }\OperatorTok{=}\NormalTok{ ko\_assignments.}\BuiltInTok{str}\NormalTok{.replace(}\StringTok{\textquotesingle{}ko:\textquotesingle{}}\NormalTok{, }\StringTok{\textquotesingle{}\textquotesingle{}}\NormalTok{).dropna()}

\CommentTok{\# Generate unique KO list per sample}
\NormalTok{unique\_kos }\OperatorTok{=}\NormalTok{ ko\_assignments.unique()}
\end{Highlighting}
\end{Shaded}

\end{document}